\documentclass[usenatbib]{mn2e}

\pdfoutput=1

\usepackage[utf8]{inputenc}
\usepackage{url}
\usepackage{fixltx2e}
\usepackage{mathptmx}
\usepackage[pdftex]{graphicx}
\usepackage{color}

\newcommand{\Mpc}{\rm\thinspace Mpc}
\newcommand{\kpc}{\rm\thinspace kpc}

\newcommand{\km}{\rm\thinspace km}

\newcommand{\cm}{\rm\thinspace cm}

\newcommand{\pcmcu}{\hbox{$\cm^{-3}\,$}}

\newcommand{\yr}{\rm\thinspace yr}
\newcommand{\Gyr}{\rm\thinspace Gyr}
\newcommand{\Myr}{\rm\thinspace Myr}
\newcommand{\s}{\rm\thinspace s}

\newcommand{\keVcmsq}{\hbox{$\keV\cm^{2}\,$}}

\newcommand{\g}{\rm\thinspace g}
\newcommand{\gpcm}{\hbox{$\g\cm^{-3}\,$}}

\newcommand{\Msun}{\hbox{$\rm\thinspace M_{\odot}$}}

\newcommand{\Msunpyr}{\hbox{$\Msun\yr^{-1}\,$}}
\newcommand{\keV}{\rm\thinspace keV}

\newcommand{\erg}{\rm\thinspace erg}

\newcommand{\ergps}{\hbox{$\erg\s^{-1}\,$}}

\newcommand{\kmps}{\hbox{$\km\s^{-1}\,$}}

\newcommand{\kmpspMpc}{\hbox{$\kmps\Mpc^{-1}$}}
\newcommand{\Zsun}{\hbox{$\thinspace \mathrm{Z}_{\odot}$}}

\newcommand{\psqcm}{\hbox{$\cm^{-2}\,$}}
\newcommand{\pcmsq}{\hbox{$\cm^{-2}\,$}}

\newcommand{\ps}{\hbox{$\s^{-1}\,$}}

\begin{document}

\title[Looking deeply into PKS 0745--191]
{Feedback, scatter and
  structure in the core of the PKS 0745--191 galaxy cluster}

\author
[J.~S. Sanders et al.]
{
  \begin{minipage}[b]{\linewidth}
    J.~S.~Sanders$^{1}$,
    A.~C. Fabian$^2$,
    J.~Hlavacek-Larrondo$^{3,4,5}$,
    H.~R.~Russell$^2$,
    G.~B.~Taylor$^6$,
    F.~Hofmann$^{1}$,
    G.~Tremblay$^7$ and
    S.~A. Walker$^2$
  \end{minipage}
  \\
  $^1$ Max-Planck-Institut für extraterrestrische Physik,
  Giessenbachstrasse 1,  85748  Garching, Germany\\
  $^2$ Institute of Astronomy, Madingley Road, Cambridge. CB3 0HA\\
  $^3$ Kavli Institute for Particle Astrophysics and Cosmology, Stanford University, 452 Lomita Mall, Stanford, CA 94305-4085, USA\\
  $^4$ Department of Physics, Stanford University, 452 Lomita Mall, Stanford, CA 94305-4085, USA\\
  $^5$ Département de Physique, Université de Montréal, C.P. 6128, Succ. Centre-Ville, Montréal, Québec H3C 3J7, Canada\\
  $^6$ Department of Physics and Astronomy, University of New Mexico,
  Albuquerque, NM 87131. USA\\
  $^7$ European Southern
  Observatory, Karl-Schwarzschild-Strasse 2, 85748 Garching, Germany\\
}
\maketitle

\begin{abstract}
  We present \emph{Chandra X-ray Observatory} observations of the core
  of the galaxy cluster PKS\,0745--191.  Its centre shows X-ray
  cavities caused by AGN feedback and cold fronts with an associated
  spiral structure. The cavity energetics imply they are powerful
  enough to compensate for cooling.  Despite the evidence for AGN feedback,
  the \emph{Chandra} and \emph{XMM}-RGS X-ray spectra are consistent
  with a few hundred solar masses per year cooling out of the X-ray
  phase, sufficient to power the emission line nebula. The coolest
  X-ray emitting gas and brightest nebula emission is offset by around
  5 kpc from the radio and X-ray nucleus. Although the cluster has a
  regular appearance, its core shows density, temperature and pressure
  deviations over the inner 100 kpc, likely associated with the cold
  fronts.  After correcting for ellipticity and projection effects, we
  estimate density fluctuations of $\sim 4$ per cent, while
  temperature, pressure and entropy have variations of $10-12$ per
  cent.  We describe a new code, \textsc{mbproj}, able to accurately
  obtain thermodynamical cluster profiles, under the assumptions of
  hydrostatic equilibrium and spherical symmetry. The forward-fitting code compares model to
  observed profiles using Markov Chain Monte Carlo and is applicable
  to surveys, operating on 1000 or fewer counts. In PKS0745 a very low
  gravitational acceleration is preferred within 40 kpc radius from the core,
  indicating a lack of hydrostatic equilibrium, deviations from
  spherical symmetry or non-thermal sources of pressure.
\end{abstract}


\section{Introduction}
The radiative cooling time of the dominant baryonic component, the
intracluster medium (ICM), in the centres of many clusters of galaxies
is much shorter than the age of the cluster. In the absence of any
heating, a cooling flow should form \citep{Fabian94} where the
material should rapidly cool out of the X-ray band at 10s to 100s
solar masses per year. Mechanical feedback by AGN in cluster cores can
energetically provide the balancing source of heat
\citep{McNamaraNulsen12}, although the tight balance suggests that the
heating is gentle and close to continuous \citep{Fabian12}.

An ideal place to study the mechanisms of feedback and the effect of
the AGN on the surrounding cluster is in the most extreme objects.
PKS\,0745--191 is the radio source located at the centre of a rich
galaxy cluster at a redshift of 0.1028. The cluster has been well
studied by X-ray observatories since the launch of \emph{Einstein}
\citep{Fabian85}. The galaxy cluster is relaxed and has a
steeply-peaked surface brightness profile. It is the X-ray brightest
at $z > 0.1$ \citep{Edge90}, with a 2 to 10 keV X-ray luminosity of
$1.6 \times 10^{45} \ergps$ and is the nearest cluster with a mass
deposition rate inferred from the X-ray surface brightness profile of
greater than $1000\Msunpyr$ \citep{Allen96}.

The central galaxy is undergoing considerable star formation. Optical
spectra suggest $\sim 50\Msunpyr$ of star formation
\citep{Johnstone87} and infrared observations using \emph{Spitzer} give
its total IR luminosity as $3.8 \times 10^{44}\ergps$, which corresponds to a
star formation rate of $17 \Msunpyr$ \citep{ODea08}.  In addition, the
cluster is a strong emitter in H$\alpha$ (L[H$\alpha$+N\textsc{ii}] =
$2.8\times 10^{42}\ergps$; \citealt{Heckman89}), in an extended
filamentary nebula, the luminosity of which is on the extreme end of
the distribution of values observed in clusters \citep{Crawford99}.
\cite{SalomeCombes03} have also detected CO(1-0) and CO(2-1)
emission lines in PKS0745, implying an H$_2$ mass of
$4\times10^9\Msun$. \cite{Hicks05} found a UV excess from the central
galaxy of $\sim 12 \times 10^{43}\ergps$. This is clearly an
interesting object where heating is not matching cooling.

The very unusual and interesting radio source in PKS0745 was studied
in detail by \cite{BaumODea91}. The radio jets appear to have been
disrupted on small scales so that the large scale structure is
amorphous.  This is seen in just a few other radio galaxies in dense
environments (e.g. 3C317 in A2052, \citealt{Zhao93}; and PKS\,1246--410
in the Centaurus cluster, \citealt{Taylor02}).

PKS0745 is one of the nearest galaxy clusters to exhibit strong
gravitational lensing \citep{AllenLens96}. The 0-th order image of a
\emph{Chandra} HETGS observation of the cluster was examined by
\cite{Hicks02}, finding a central temperature of $4-5\keV$. The
properties of the ICM in PKS0745 have been measured to beyond the
virial radius \citep{George09} using \emph{Suzaku}, showing a
considerable flattening of the entropy profile. \cite{Walker12} have
improved on this analysis with the aid of further observations to
better understand the X-ray background contribution.  They find that
at radii beyond $1.9\Mpc$ the cluster is not in hydrostatic
equilibrium or there is significant non-thermal pressure support.

At the redshift of the cluster, 1 arcsec on the sky corresponds to
1.9~kpc, assuming that $H_0 = 70 \kmpspMpc$. In this paper, the
relative Solar abundances of \cite{AndersGrevesse89} were used. North
is to the top and east is to the left in images, unless otherwise
indicated. The Galactic Hydrogen column density towards PKS0745
is high, around $4.2\times10^{21}\pcmsq$, weighting the nearest 0.695
deg pixels in the Leiden/Argentine/Bonn (LAB) survey
\citep{Kalberla05}. However, these pixels show large variation with a
standard deviation of $4 \times 10^{20}\pcmsq$.

\section{Chandra images}
\label{sect:images}
\begin{figure}
  \centering
  \includegraphics[width=\columnwidth]{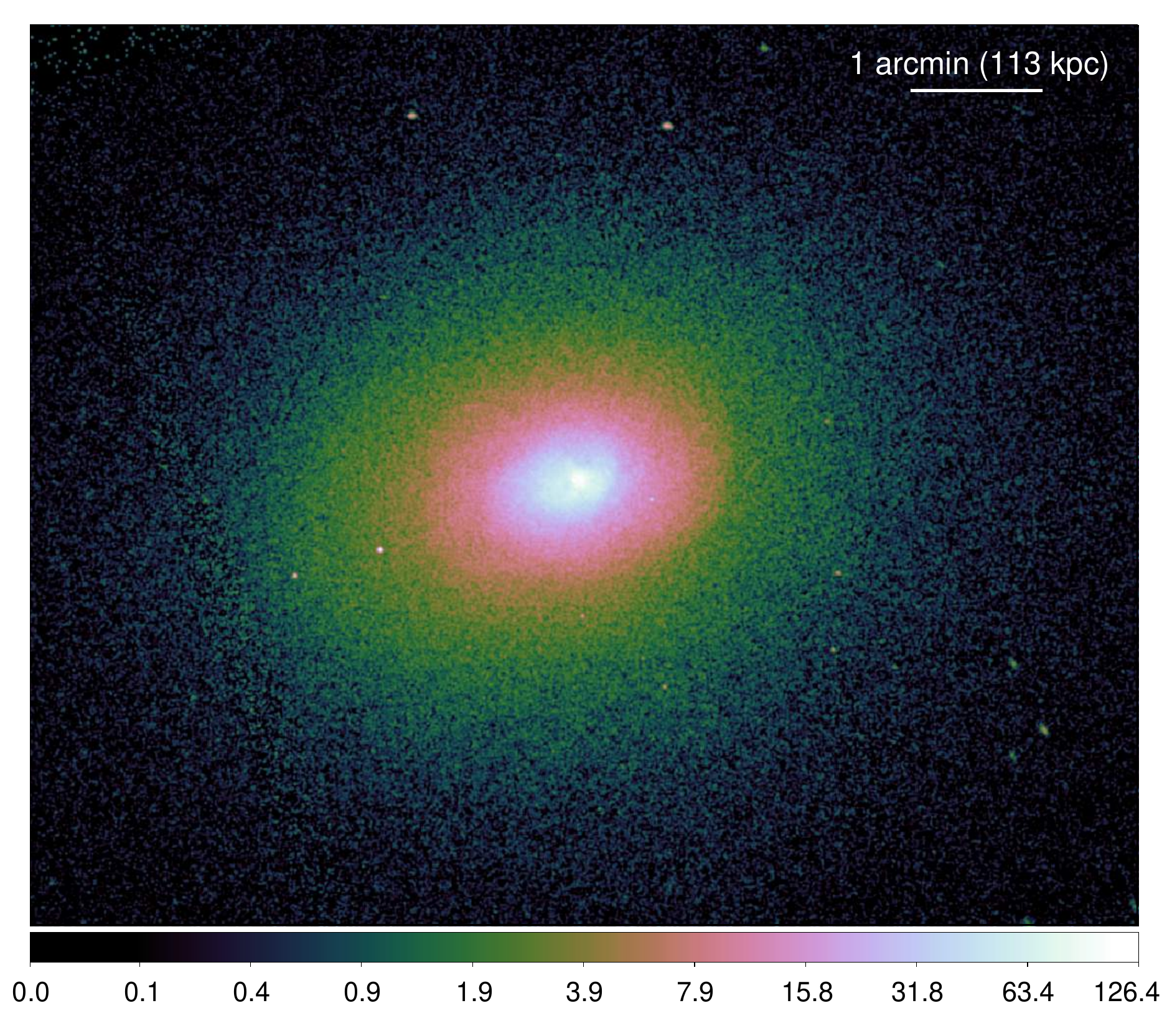}
  \caption{Exposure-corrected and background-subtracted 0.5
    to 7 keV image using 0.492 arcsec pixels, smoothed with a Gaussian
    with $\sigma=1$ pixel. The units are $10^{-8}$ photon
    $\ps\psqcm$.}
  \label{fig:img}
\end{figure}

\begin{figure}
  \centering
  \includegraphics[width=\columnwidth]{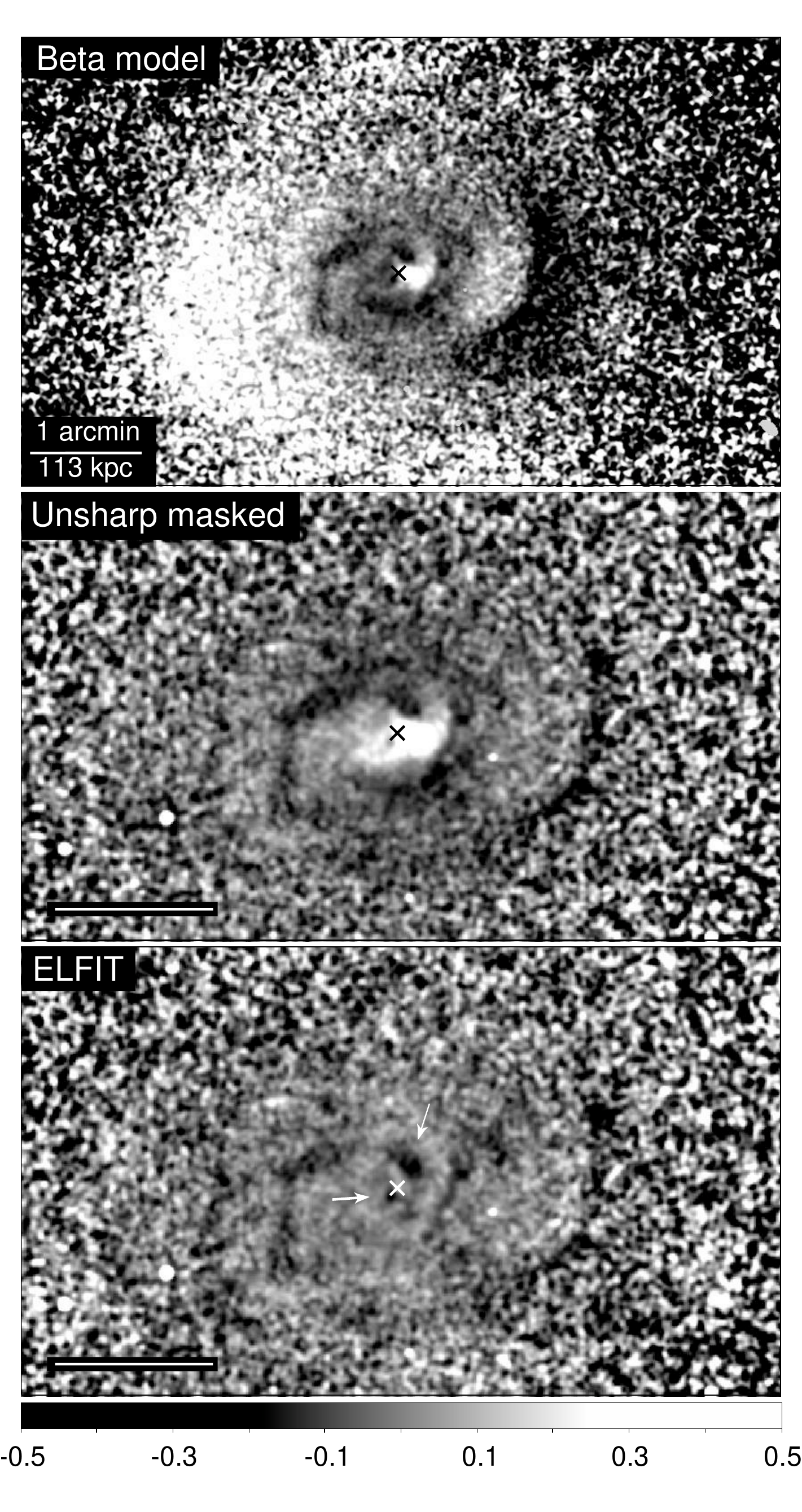}
  \caption{(Top panel) Fractional residuals of $\beta$ model fit to
    0.5 to 7 keV exposure-corrected and background-subtracted
    data. (Centre panel) Unsharp masked image, showing fractional
    residuals between data smoothed by a Gaussian of $\sigma=2$ and 16
    pixels (1 and 8 arcsec). (Bottom panel) Fractional difference
    between data (smoothed by 2 pixels) and \textsc{elfit}
    multi-elliptical model. Note that the lower two panels are zoomed
    in compared to the top panel. The white arrows in the bottom panel
    show the location of two depressions. `$\times$' marks the
    location of the central radio source.}
  \label{fig:resid}
\end{figure}

We examined two ACIS-S \emph{Chandra} observations of PKS0745
with observation identifiers of 12881 (exposure time of 118.1~ks) and
2427 (exposure time of 17.9~ks). We did not examine observation 507
which contained extensive flares. We reprocessed the observations to
filter background events using VFAINT mode. To remove flares we used
an iterative $\sigma$-clipping algorithm on lightcurves from the CCD
S1, which is back-illuminated like ACIS S3, in the 2.5 to 7 keV
band. No additional flares were seen in a higher energy band. The
total cleaned exposure of the observations was 135.5~ks.  The total
exposure-corrected and background-subtracted image of the
cluster is shown in Fig.~\ref{fig:img}.  For the background
subtraction we used blank-sky-background event files. Separate
backgrounds were used for each observation with the the exposure
time in each CCD adjusted so that the count rate in the 9 to 12 keV
band matched the respective observation. The background event
files were reprojected to match the observations. The background images for
each observation and CCD were divided by the ratio of the background
to observation exposure times, before subtracting from the observation
image.  A monochromatic exposure map at 1.5~keV was then used
for exposure correction.

Due to the bright central peak, it is difficult to look for structure
in the X-ray image. The features are seen more easily by looking at
the residuals of a model fit or by using unsharp-masking. The top
panel of Fig.~\ref{fig:resid} shows the residuals from an elliptical
$\beta$ model fit to the surface brightness.  The centre panel shows a
closer view of the central region, but using unsharp masking to remove
the larger scale emission. The bottom panel shows the residuals to an
\textsc{elfit} model made up by logarithmically interpolating between 20
ellipses fitted to logarithmically spaced contours of surface
brightness \citep{SandersAWM712}.  In all residuals, a spiral
morphology can be seen. In addition, there is a sharp drop in surface
brightness $\sim 1$ arcmin to the west of the core, some of which is
modelled out by the \textsc{elfit} procedure.  The centre of the cluster
contains two central depressions in surface brightness to the
north-west and south-east (see white arrows in bottom panel). The
depressions are around 20 per cent fainter than the surrounding
material.  There are other possible features in the central region,
but many of these appear to be related to the swirling spiral
feature. The negative spiral residuals are around 10 per cent in
magnitude.

\begin{figure}
  \centering
  \includegraphics[width=0.85\columnwidth]{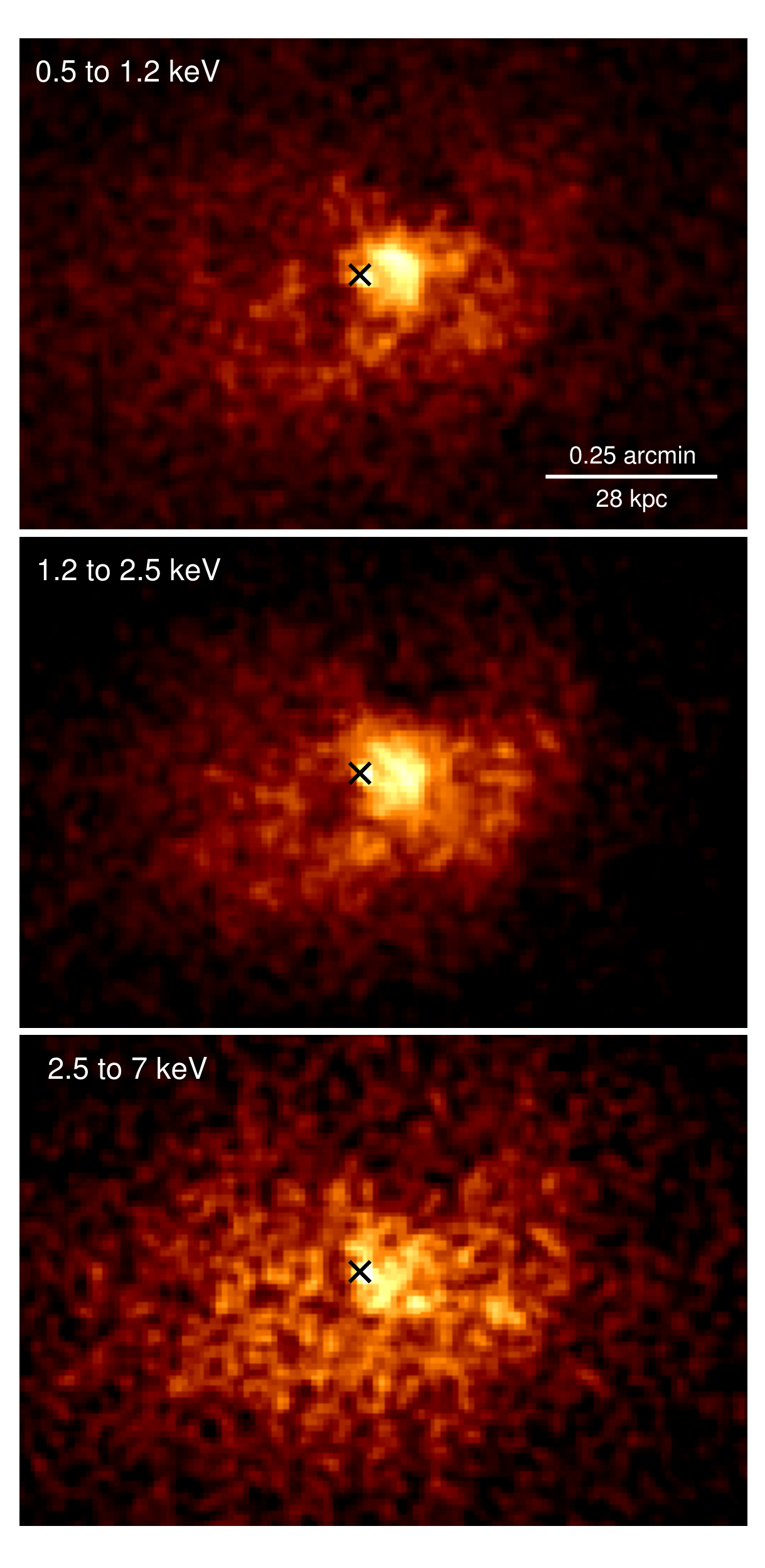}
  \caption{Unsharp-masked images in the different spectral
    bands. These images have the original image smoothed by a Gaussian
    with $\sigma=1$ pixel, subtracting 0.5 times the image smoothed by
    8 pixels. `$\times$' marks the location of the central radio
    source.}
  \label{fig:bands}
\end{figure}

Fig.~\ref{fig:bands} shows the centre of the cluster in soft, medium
and hard bands. The image shows that there is soft X-ray emission in
the rims of the surface brightness depressions. In addition, there is
an extended bright region of soft X-ray emission a few arcsec to the
west of the central nucleus, seen as a point source. We examine this
in more detail in Section \ref{sect:nucleus}.

\begin{figure}
  \centering
  \includegraphics[width=\columnwidth]{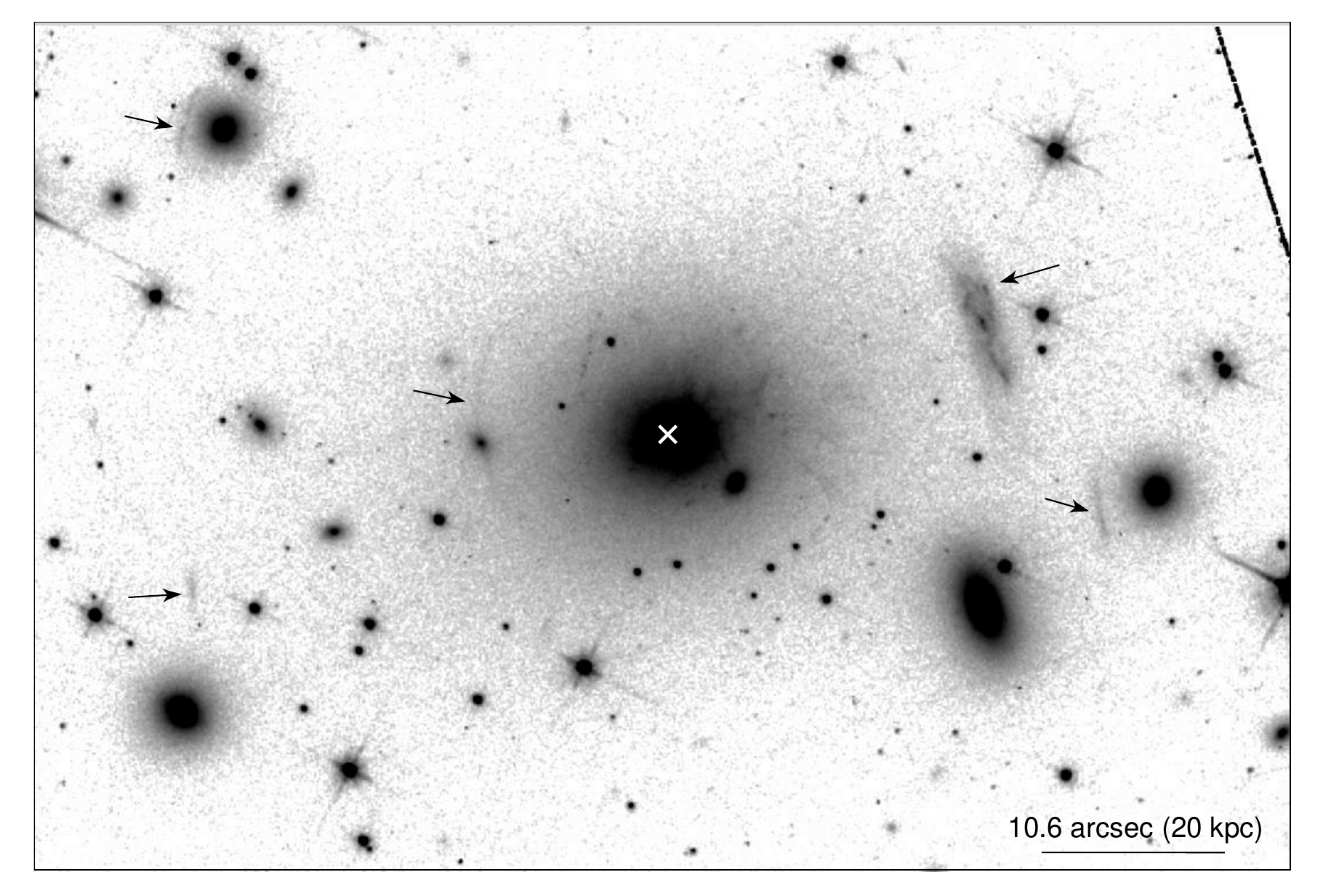}
  \caption{\emph{HST} F814W image of the centre of the
    cluster. Positions of possible strong lensing arcs are marked by
    arrows and `$\times$' shows the central radio source
      position. The image was created from WFPC2 datasets U59N0101R,
      U59N0102R and U59N0103R.}
  \label{fig:hst}
\end{figure}

\section{Nuclear region}
\label{sect:nucleus}

The brightest central galaxy (BCG) was observed by \emph{HST} in 1999
using the WFPC2 instrument with F184W and F55W filters, with exposure
times of $3\times600$ and $3\times700$~s, respectively. At the
redshift of PKS0745, the F814W band contains the
H$\alpha$+N\textsc{ii} emission lines. In Fig.~\ref{fig:hst} we show
the F814W image of the centre of the cluster. A number of strong
lensing arcs can be seen in the image.  We have marked possible arcs
with arrows. The strongest arc is 18 arcsec to the west-north-west
(WNW) of the core. Another arc 12 arcsec to the east is also clearly
seen. \cite{AllenLens96} noted three arcs, which were the strong arc
to the WNW, the other south of that, plus another further south. We do
not see this final arc, but see three further candidates to the
east. An emission line nebula is present in the inner 20 kpc, but is
difficult to see against the continuum of the galaxy in this
image.

\begin{figure}
  \centering
  \includegraphics[width=\columnwidth]{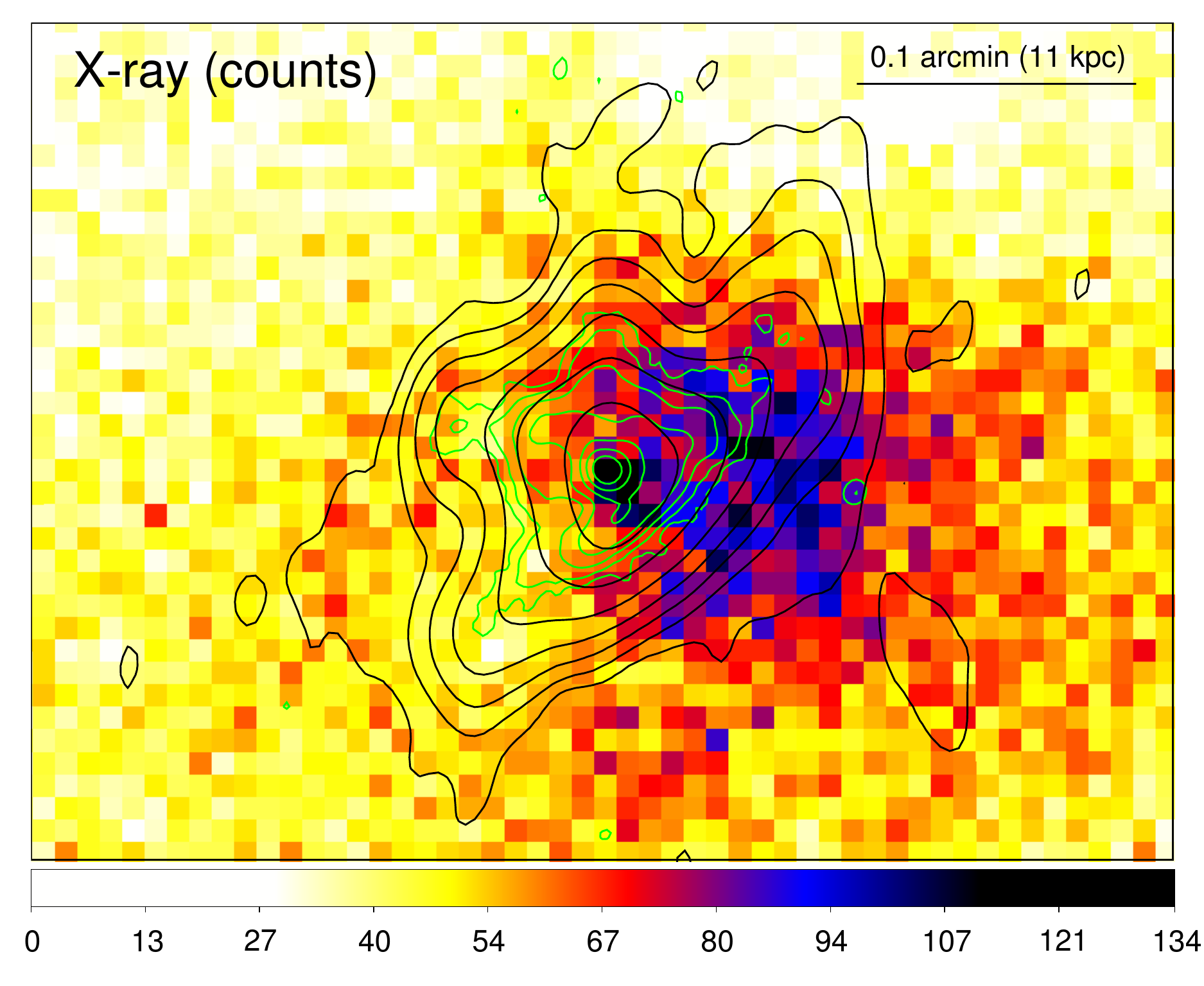}
  \includegraphics[width=\columnwidth]{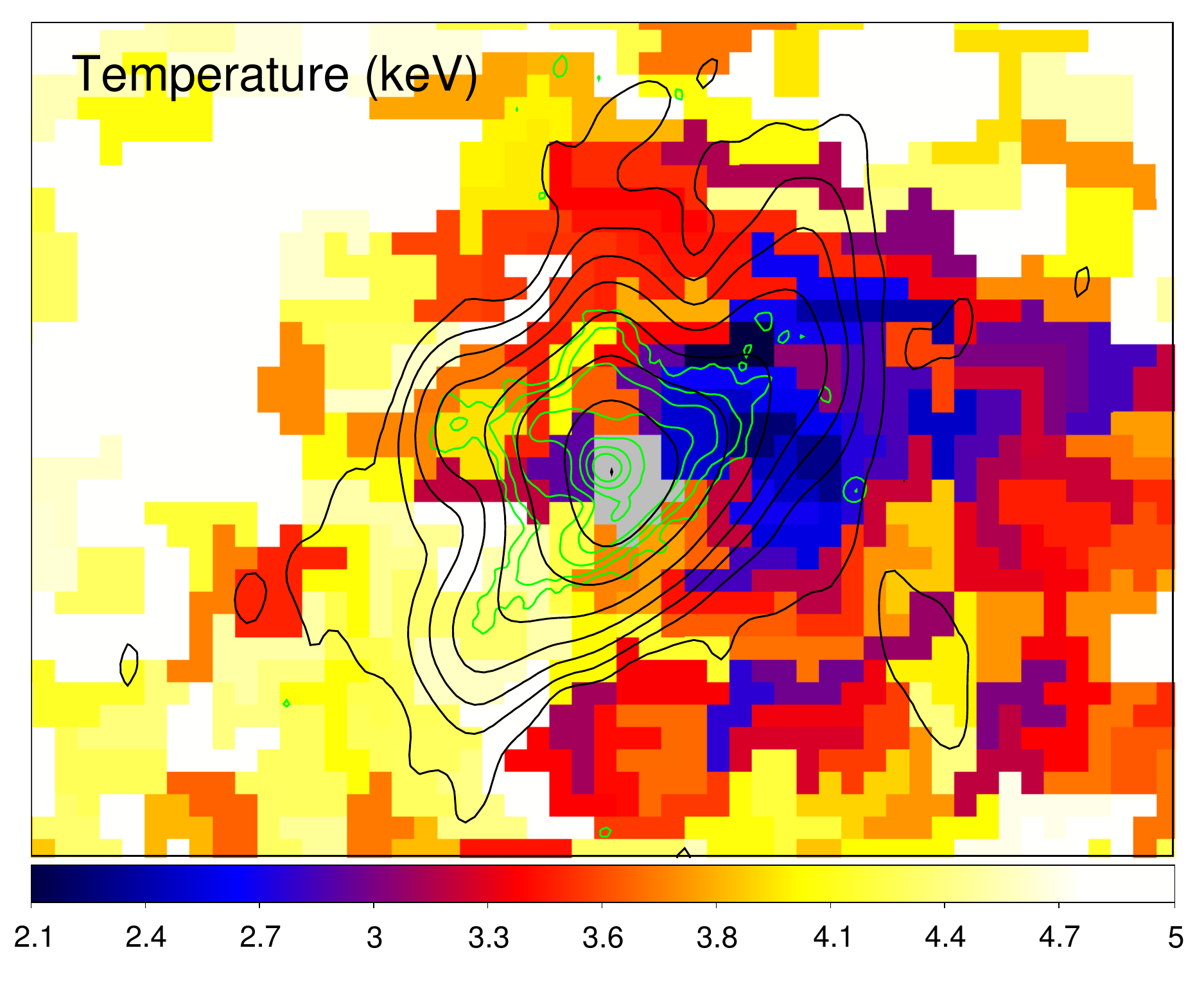}
  \includegraphics[width=\columnwidth]{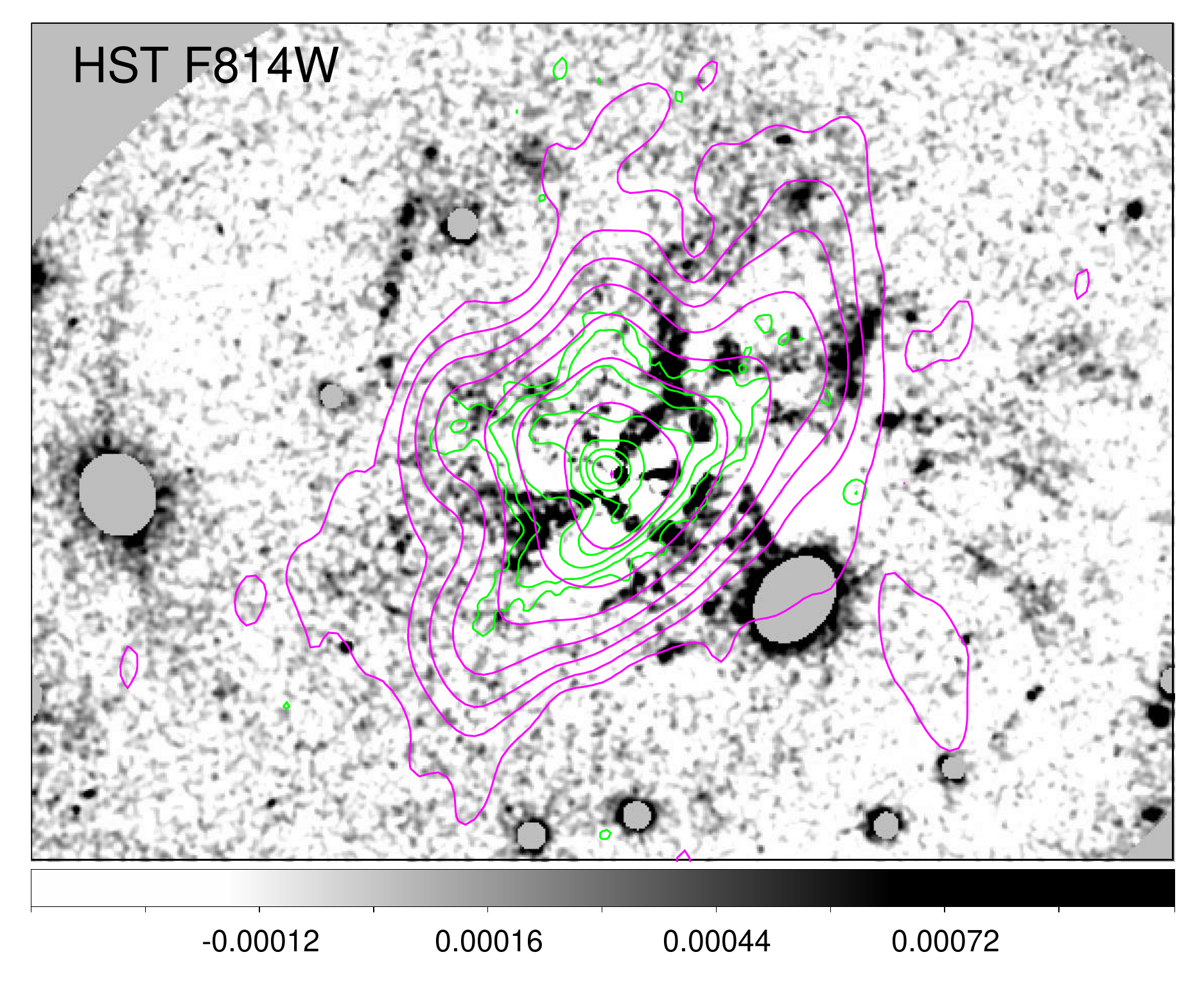}
  \caption{(Top panel) X-ray count image of the nuclear
    region of the cluster between 0.5 and 7 keV. The smaller contours
    show 8.4~GHz VLA-detected radio emission \protect\citep{Taylor94},
    logarithmically spaced between $5 \times 10^{-5}$ and 0.05
    Jy~beam$^{-1}$. The larger contours show 1.4 GHz emission with
    8 contours between 0.0003 and 0.66 Jy~beam$^{-1}$.  (Centre panel) X-ray
    temperature map in keV from spectral fitting of regions with a
    signal to noise of 20 ($\sim 400$ counts). (Bottom panel)
    \emph{HST} F814W image after subtraction of a smooth model. The grey
    regions are excluded point sources or galaxies, or lie outside the smooth model.}
  \label{fig:nuclimg}
\end{figure}

\begin{figure}
  \centering
  \includegraphics[width=\columnwidth]{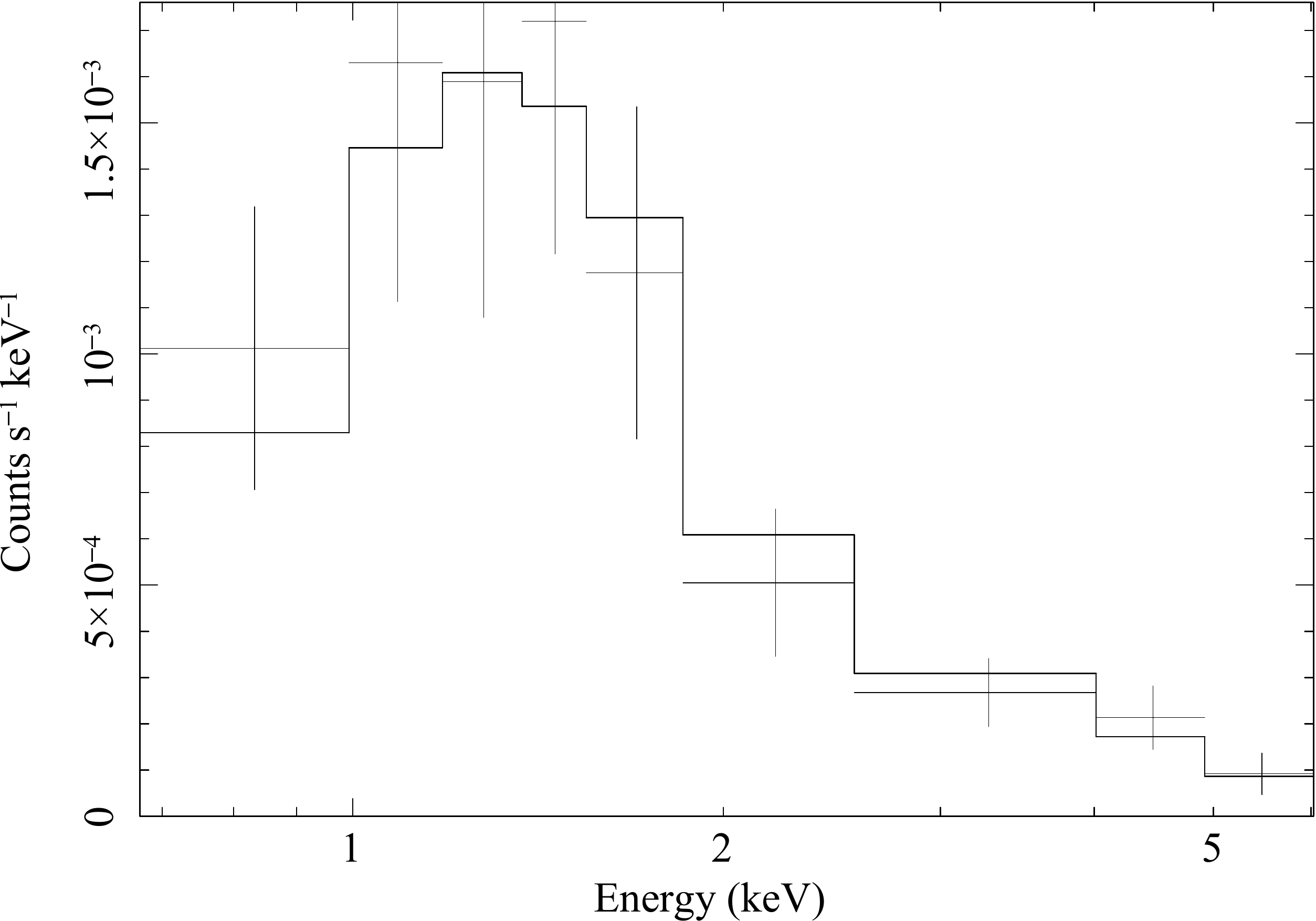}
  \caption{Powerlaw fit to the spectrum of the nucleus. Assuming
    Galactic absorption, the photon index of $1.8 \pm 0.2$ and its
    luminosity is $(6.2 \pm 1.2) \times 10^{41}\ergps$ between 2 and
    10 keV. The spectrum is rebinned for display to have a minimum
    signal to noise ratio of 3 in each spectral bin.}
  \label{fig:nuclspec}
\end{figure}

An X-ray image of the region around the nucleus can be seen in the top
panel of Fig.~\ref{fig:nuclimg}. The point source associated with the
nucleus is observed to the east of a brighter region of
emission. Fig.~\ref{fig:nuclspec} shows the spectrum of the nucleus
extracted from a 1 arcsec radius region, using a background region
from 1 to 2.4 arcsec radius. We fit the spectrum between 0.5 and 7
keV, minimising the C-statistic. It can be fit with either a powerlaw
or a thermal model.  Assuming that there is no additional absorption
above Galactic values (using the value of $4.2 \times 10^{21}\pcmsq$;
\citealt{Kalberla05}), the best fitting temperature is 5.1~keV. The
more physical powerlaw model gives a photon index of $1.8 \pm 0.2$ and
has a 2 to 10~keV luminosity of $(6.2 \pm 1.2) \times
10^{41}\ergps$. \cite{HlavacekLarrono11} found a luminosity of $(7.5
\pm 3.0) \times 10^{41} \ergps$ by spectral fitting a powerlaw to the
previous short \emph{Chandra} observation, assuming no additional
absorption.

As part of this programme we performed observations between 1-2 GHz
with the newly upgraded Very Large Array (VLA).  The observations
lasted for five hours on 21 October 2012 and consisted of 16 spectral
windows of 64 MHz each covering the entire 1-2 GHz band.  In
Fig.~\ref{fig:nuclimg} we show contours of the 1376 MHz radio
emission, derived from a single spectral window. The data were
calibrated and imaged in AIPS using standard procedures. Results from
a full multi-frequency synthesis using the entire band will be
presented in a future paper.

We constructed an X-ray temperature map using regions with a signal to
noise ratio of 20 (around 400 counts per bin). The regions were
selected using the contour binning algorithm of \cite{SandersBin06},
using a maximum ratio between the length and width of the bins of 2,
and creating bins by following the surface brightness on a 0.5 to 7
keV image smoothed with a kernel adjusted to have a signal to noise
ratio of 15 (225 counts). Spectra were extracted from each of the
observations and combined. We created response and ancillary response
matrices for each bin, weighting the response regions by the number of
counts in the 0.5 to 7 keV band. Background spectra were extracted
from standard background event files, reprojected to match the
foreground observations. The background event files exposure times
were adjusted to match the count rate in the foreground observations
in the 9 to 12 keV band. The background spectra for the different
observations were added, throwing away photons from the shorter
observation background spectra to maintain the ratio of effective
exposure time between the observations foregrounds and
backgrounds. When spectral fitting, the Galactic absorption was fixed
at $4.2\times10^{21} \pcmsq$ and the metallicity was frozen at
$0.44\Zsun$ (an average value of the centre taken from maps created
with higher signal to noise per bin). We used the \textsc{apec}
thermal spectral model \citep{SmithApec01} with \textsc{phabs}
photoelectric absorption \citep{BalucinskaChurchPhabs92}, fitting
between 0.5 and 7 keV.

It can be seen that the coolest X-ray emitting gas lies a few kpc to
the west of the nucleus (Fig.~\ref{fig:nuclimg} centre panel). The
minimum projected temperature drops to around 2.1~keV. The radio
source at 1.4 and 8.4~GHz shows a complex morphology. The coolest X-ray
emitting material lies close to the the western extension of the
radio source, but there is little cool material in other directions. This
coolest material does not have a significantly different metallicity
from nearby gas ($0.38 \pm 0.05 \Zsun$). If we allow the metallicity
to be free when fitting the map spectra, we do not see any regions
with anomalously low or high metallicity.

Fig.~\ref{fig:hst} showed an \emph{HST} image of the cluster (see
Section \ref{sect:images}). In Fig.~\ref{fig:nuclimg} (bottom panel)
we reveal the H$\alpha$ emitting filaments by subtracting a smooth
model. This was an \textsc{elfit} elliptical model fitted to 20
surface brightness contours, excluding other bright sources. The
filaments to the north-west appear to follow the 1.4~GHz
contours. This is where the bulk of the emission is concentrated and
is coincident with the coolest X-ray emitting gas. However, other
filaments (e.g. the radial filaments to the west) are not coincident
with currently detected radio emission.

\section{Central spectra}

\begin{table}
  \caption{Best fitting values in spectral fits to the \emph{Chandra}
    data extracted from the central 1.0 arcmin region, where the
    cooling time is less than 7.7~Gyr.  The results are for models
    from one to three thermal components ($\textsc{apec}$), with
    photoelectric absorption applied ($\textsc{phabs}$). We also
    examine a cooling flow model \textsc{mkcflow}, which assumes the
    gas is radiatively cooling between two temperatures at a certain
    mass deposition rate ($\dot{M}$).  Shown are the absorbing column
    densities ($n_\mathrm{H}$), temperatures ($T$), metallicities
    ($Z$) and model normalisations (Norm; measured in \textsc{xspec}
    normalisation units).  In these fits we assumed that each
    component had the same metallicity.  For the cooling flow model,
    we tied the temperature of the \textsc{apec} component to the
    upper temperature of the \textsc{mkcflow} model.}
  \centering
  \begin{tabular}{llc}
    \hline
    Model                            & Parameter                         & Value                    \\ \hline
    \textsc{phabs(apec)}             & $n_\mathrm{H}$ ($10^{22} \pcmsq$) & $0.372 \pm 0.002$        \\
                                     & $Z$ ($Z_\odot$)                   & $0.40 \pm 0.01$          \\
                                     & $T$ (keV)                         & $6.38 \pm 0.05$          \\
                                     & Norm ($10^{-4}$)                  & $377 \pm 1$              \\
                                     & $\chi^2_\nu$                      & $1.32 = 579/440$         \\
    \textsc{phabs(apec+apec)}        & $n_\mathrm{H}$ ($10^{22} \pcmsq$) & $0.398 \pm 0.004$        \\
                                     & $Z$ ($Z_\odot$)                   & $0.40 \pm 0.01$          \\
                                     & $T_1$ (keV)                       & $6.24 \pm 0.05$          \\
                                     & $T_2$ (keV)                       & $0.67 \pm 0.04$          \\
                                     & Norm 1 ($10^{-4}$)                & $381 \pm 2$              \\
                                     & Norm 2 ($10^{-4}$)                & $7.7 \pm 1.3$            \\
                                     & $\chi^2_\nu$                      & $1.19 = 522 / 438$       \\
    \textsc{phabs(apec+apec+apec)}   & $n_\mathrm{H}$ ($10^{22} \pcmsq$) & $0.409 \pm 0.004$        \\
                                     & $Z$ ($Z_\odot$)                   & $0.43 \pm 0.01$          \\
                                     & $T_1$ (keV)                       & $10.9 \pm 0.8$           \\
                                     & $T_2$ (keV)                       & $4.8 \pm 0.2$            \\
                                     & $T_3$ (keV)                       & $0.64 \pm 0.03$          \\
                                     & Norm 1 ($10^{-4}$)                & $140 \pm 15$             \\
                                     & Norm 2 ($10^{-4}$)                & $246 \pm 16$             \\
                                     & Norm 3 ($10^{-4}$)                & $9.7 \pm 1.2$            \\
                                     & $\chi^2_\nu$                      & $1.11 = 485 / 436$       \\
    \textsc{phabs(apec+mkcflow)}     & $n_\mathrm{H}$ ($10^{22} \pcmsq$) & $0.393 \pm 0.003$        \\
                                     & $Z$ ($Z_\odot$)                   & $0.40 \pm 0.01$          \\
                                     & $T_\mathrm{high}$ (keV)           & $6.65 \pm 0.06$          \\
                                     & $T_\mathrm{low}$ (keV)            & $0.17 \pm 0.03$          \\
                                     & Norm ($10^{-4}$)                  & $350 \pm 5$              \\
                                     & $\dot{M}$ ($M_\odot$ yr$^{-1}$)   & $112 \pm 20$             \\
                                     & $\chi^2_\nu$                      & $1.19 = 522 / 438$       \\
    \textsc{phabs($5\times$mkcflow)} & $n_\mathrm{H}$ ($10^{22} \pcmsq$) & $0.446 \pm 0.005$        \\
                                     & $Z$ ($Z_\odot$)                   & $0.41 \pm 0.01$          \\
                                     & $\dot{M}(T)$                      & See Fig. \ref{fig:mdots} \\
                                     & $\chi^2_\nu$                      & $1.24 = 540 / 437$       \\
    \textsc{phabs($9\times$mkcflow)} & $n_\mathrm{H}$ ($10^{22} \pcmsq$) & $0.438 \pm 0.008$        \\
                                     & $Z$ ($Z_\odot$)                   & $0.43 \pm 0.01$          \\
                                     & $\dot{M}(T)$                      & See Fig. \ref{fig:mdots} \\
                                     & $\chi^2_\nu$                      & $1.10 = 477 / 433$       \\ \hline
  \end{tabular}
  \label{tab:speccentre}
\end{table}

We extracted the \emph{Chandra} spectrum from within a radius of
1.0~arcmin, where the mean radiative cooling time is less than 7.7~Gyr
(see Section \ref{sect:radial}), the time since $z=1$. We fitted this
spectrum between 0.5 and 7 keV by minimising the $\chi^2$ statistic in
\textsc{xspec} \citep{ArnaudXspec}, after grouping it to have a
minimum of 20 counts per spectral bin.  Models (listed in Table
\ref{tab:speccentre}) with 1, 2 and 3 \textsc{apec} thermal components
were fitted, assuming that the components have the same metallicity
and are absorbed by the same \textsc{phabs} Galactic photoelectric
absorption. We also fitted the spectrum with a model with a single
thermal component plus a cooling flow model \citep{Fabian94}, which
assumes that the plasma is radiatively cooling from an upper
temperature to a lower temperature at a certain rate in $\Msunpyr$. In
this model, we use the same temperature for the thermal component as
the upper temperature of the cooling flow model and allow the lower
temperature to be free. All of these models give similar values for
the metallicity ($\sim 0.4 \Zsun$) and the absorbing column density
($\sim 0.4 \times 10^{22} \pcmsq$).

\begin{figure}
  \includegraphics[width=\columnwidth]{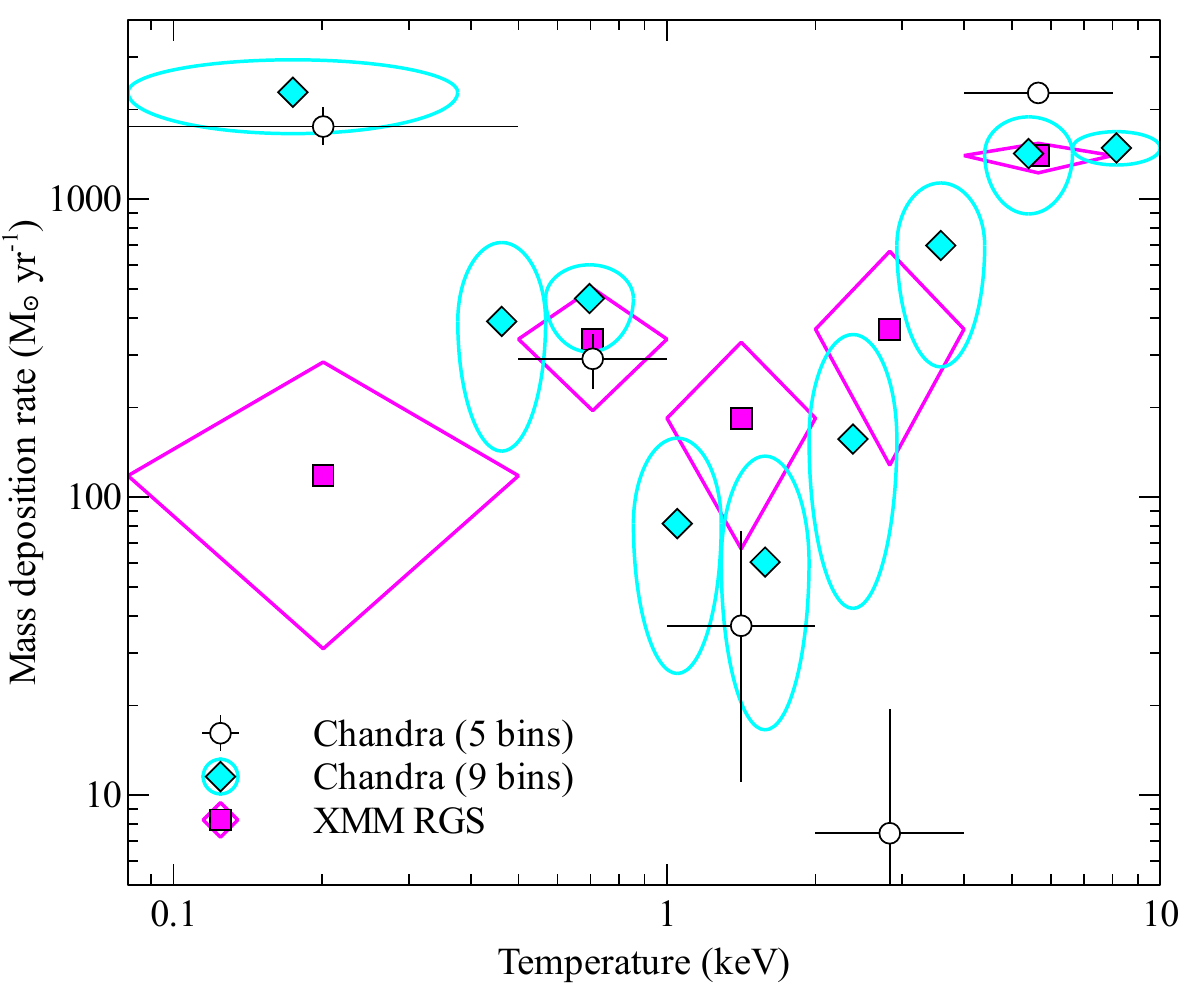}
  \caption{Central temperature distribution, parametrised as the rate
    of mass deposition as a function of temperature. The results are
    for the \emph{Chandra} spectrum from the inner arcmin radius and
    the \emph{XMM}-RGS spectra. For \emph{Chandra} we used models with
    5 (circles) and 9 (diamonds) temperature bins and for \emph{XMM}
    we used 5 bins (squares). The parameter values and uncertainties were
    obtained with an MCMC analysis.}
  \label{fig:mdots}
\end{figure}

The models support a wide range in temperature within the examined
region. We can examine the temperature distribution in more detail by
fitting the data with multiple cooling flow models in consecutive
sets of temperature intervals. This model parametrises the
distribution of gas temperature in terms of the rate of matter which
would need to be radiatively cooling in particular temperature bands
to give rise to the spectrum observed. If all matter was radiatively
cooling, this would give a constant value as a function of
temperature. Fig.~\ref{fig:mdots} shows the distribution obtained with
either 5 or 9 different ranges in temperature. In these fits, each
component was assumed to have the same metallicity and was absorbed by
the same column density (shown in Table \ref{tab:speccentre}). The
error bars were obtained from the output chain of a Markov Chain Monte
Carlo (MCMC) analysis of the spectral fit, using \textsc{emcee}
\citep{ForemanMackey12} with 200 walkers and a 2000 iterations after a
burn period of 500 iterations\footnote{Our code to use \textsc{emcee}
  with \textsc{xspec} is available at
  \url{https://github.com/jeremysanders/xspec_emcee}.}. \textsc{emcee}
uses the affine invariant MCMC sampler of \cite{Goodman10}. The
average acceptance fraction was 43 per cent and the mean
autocorrelation time in the chain was 24 iterations, indicating that
our results had converged.

\begin{table}
  \caption{Spectral fitting results for XMM RGS data. For details on the values shown, see Table \ref{tab:speccentre}.}
  \centering
  \begin{tabular}{llc}
    \hline
    Model                            & Parameter                         & Value                    \\ \hline
    \textsc{phabs(apec+apec)}        & $n_\mathrm{H}$ ($10^{22} \pcmsq$) & $0.40 \pm 0.02$          \\
                                     & $Z$ ($Z_\odot$)                   & 0.40 (fixed)             \\
                                     & $T_1$ (keV)                       & $4.4_{-0.4}^{+0.5}$      \\
                                     & $T_2$ (keV)                       & $0.84 \pm 0.08$          \\
                                     & Norm 1 ($10^{-4}$)                & $290 \pm 9$              \\
                                     & Norm 2 ($10^{-4}$)                & $11 \pm 4$               \\
                                     & C-stat                            & $4504$                   \\
    \textsc{phabs(apec+mkcflow)}     & $n_\mathrm{H}$ ($10^{22} \pcmsq$) & $0.39 \pm 0.02$          \\
                                     & $Z$ ($Z_\odot$)                   & 0.40 (fixed)             \\
                                     & $T_\mathrm{high}$ (keV)           & $5.9_{-0.8}^{+1.2}$      \\
                                     & $T_\mathrm{low}$ (keV)            & $0.49_{-0.19}^{+0.14}$   \\
                                     & Norm ($10^{-4}$)                  & $220 \pm 26$             \\
                                     & $\dot{M}$ ($M_\odot$ yr$^{-1}$)   & $267_{-85}^{+94}$        \\
                                     & C-stat                            & $4505$                   \\
    \textsc{phabs($5\times$mkcflow)} & $n_\mathrm{H}$ ($10^{22} \pcmsq$) & $0.40 \pm 0.02$          \\
                                     & $Z$ ($Z_\odot$)                   & 0.40 (fixed)             \\
                                     & $\dot{M}(T)$                      & See Fig. \ref{fig:mdots} \\
                                     & C-stat                            & $4509$                   \\ \hline
  \end{tabular}
  \label{tab:xmmrgs}
\end{table}

The cluster was also observed by \emph{XMM-Newton} for 28.3~ks using
its RGS instruments in observation 0105870101. We extracted spectra
using a cross-dispersion range containing 90 per cent of the
  PSF width (equivalent to a strip approximately 50 arcsec wide across
  the cluster) and 90 per cent of the pulse-height
distribution. Spectra were extracted in wavelength space. After
removing flares using a cut of 0.2 counts per second on CCD 9, with
flag values of 8 or 16 and an absolute cross-dispersion angle of less
than $1.5\times10^{-4}$, we obtained an exposure time of 21.6~ks. We
used a background region using the spatial region beyond 98 per cent
of the cross-dispersion PSF. Data for the two detectors were merged
using \textsc{rgscombine}.

The spectra were jointly fit, minimising the C-statistic in
\textsc{xspec}. We fit the first order spectrum between $7$ and
$20${\AA} and the second order spectrum between $7$ and $17${\AA}. Due
to the relatively poor quality of the spectra, we fixed the
metallicity when fitting to $0.40 \Zsun$, the value obtained in the
\emph{Chandra} analyses. We show the results for our spectral fitting,
using some of the models used for the \emph{Chandra} spectral fitting,
in Table \ref{tab:xmmrgs}.

A two-component thermal model obtains a lower temperature component at
$\sim 0.8$~keV, as found from fitting the \emph{Chandra} spectra. The
spectra appear consistent with around $270 \Msunpyr$ of cooling from 6
to 0.5 keV temperature. This value is larger than that obtained
from \emph{Chandra}, but if the $\dot{M}$ as a function of temperature
is examined (Fig.~\ref{fig:mdots}), the \emph{XMM} and \emph{Chandra}
values are consistent over much of that range, except at the lowest
temperatures, where results from CCD spectra are unlikely to be
accurate. It is likely that the different overall value is due to the
different temperature sensitivities of the two instruments.

\section{Central spectral maps}
\label{sect:centmaps}

\begin{figure}
  \includegraphics[width=0.99\columnwidth]{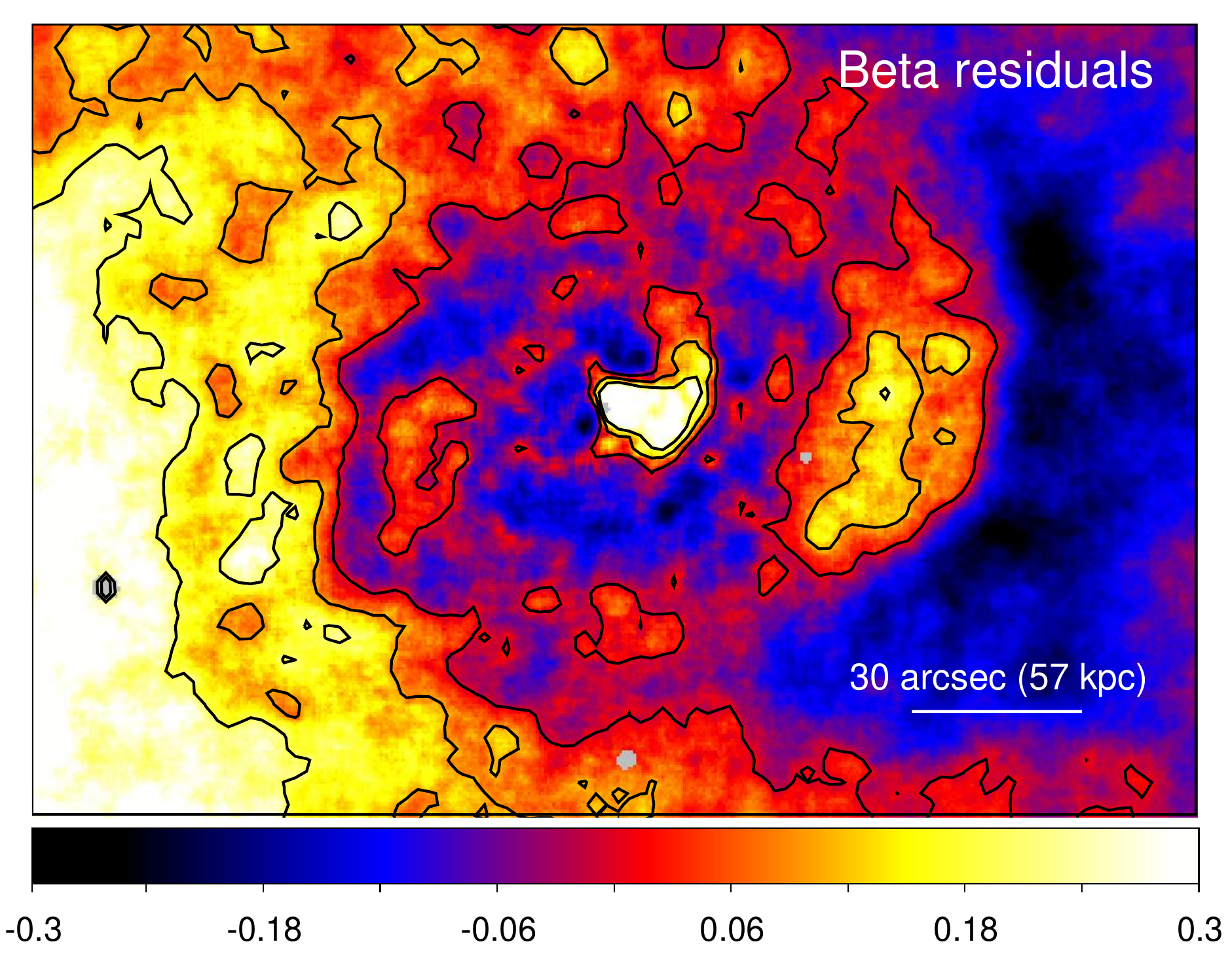}
  \includegraphics[width=0.99\columnwidth]{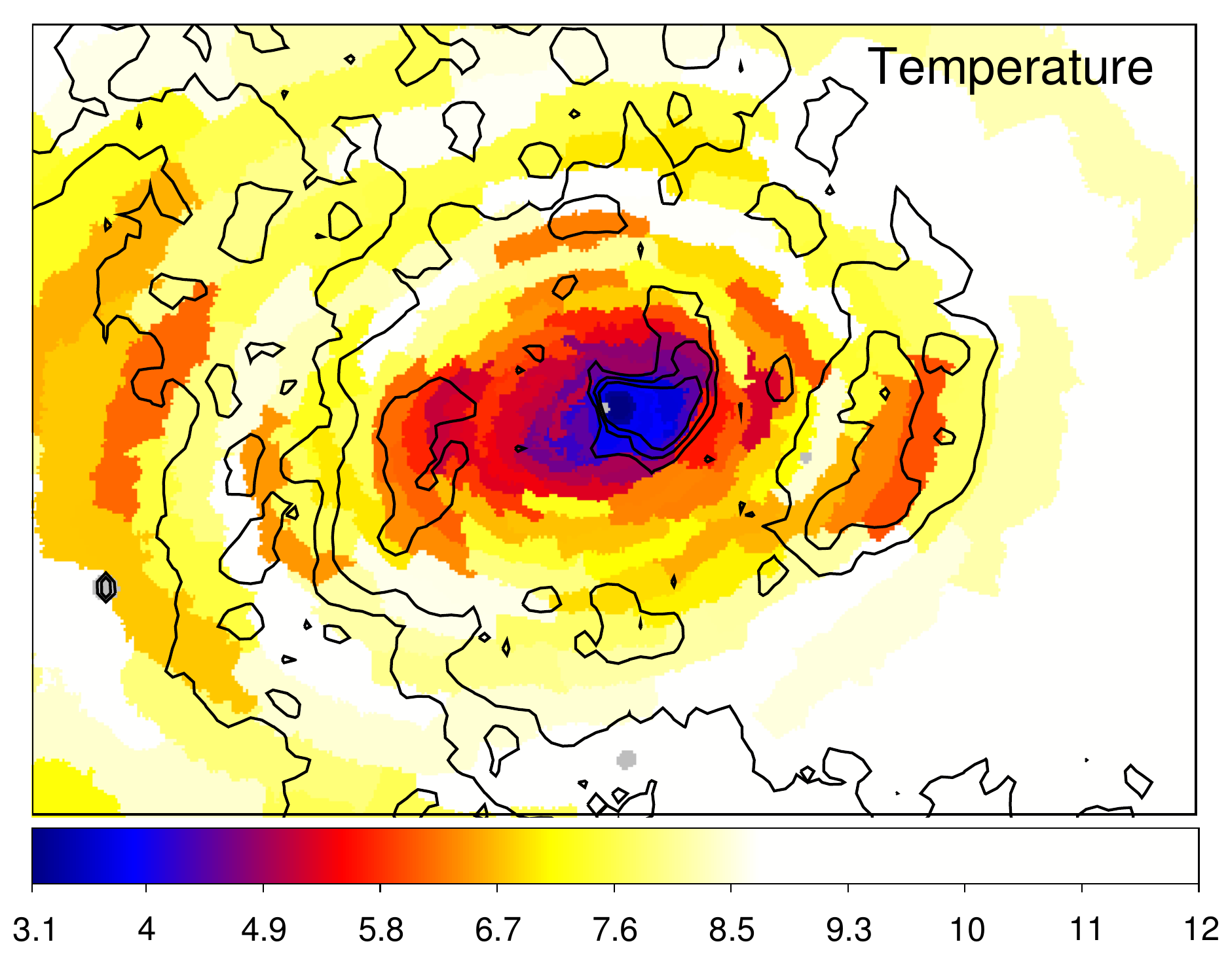}
  \includegraphics[width=0.99\columnwidth]{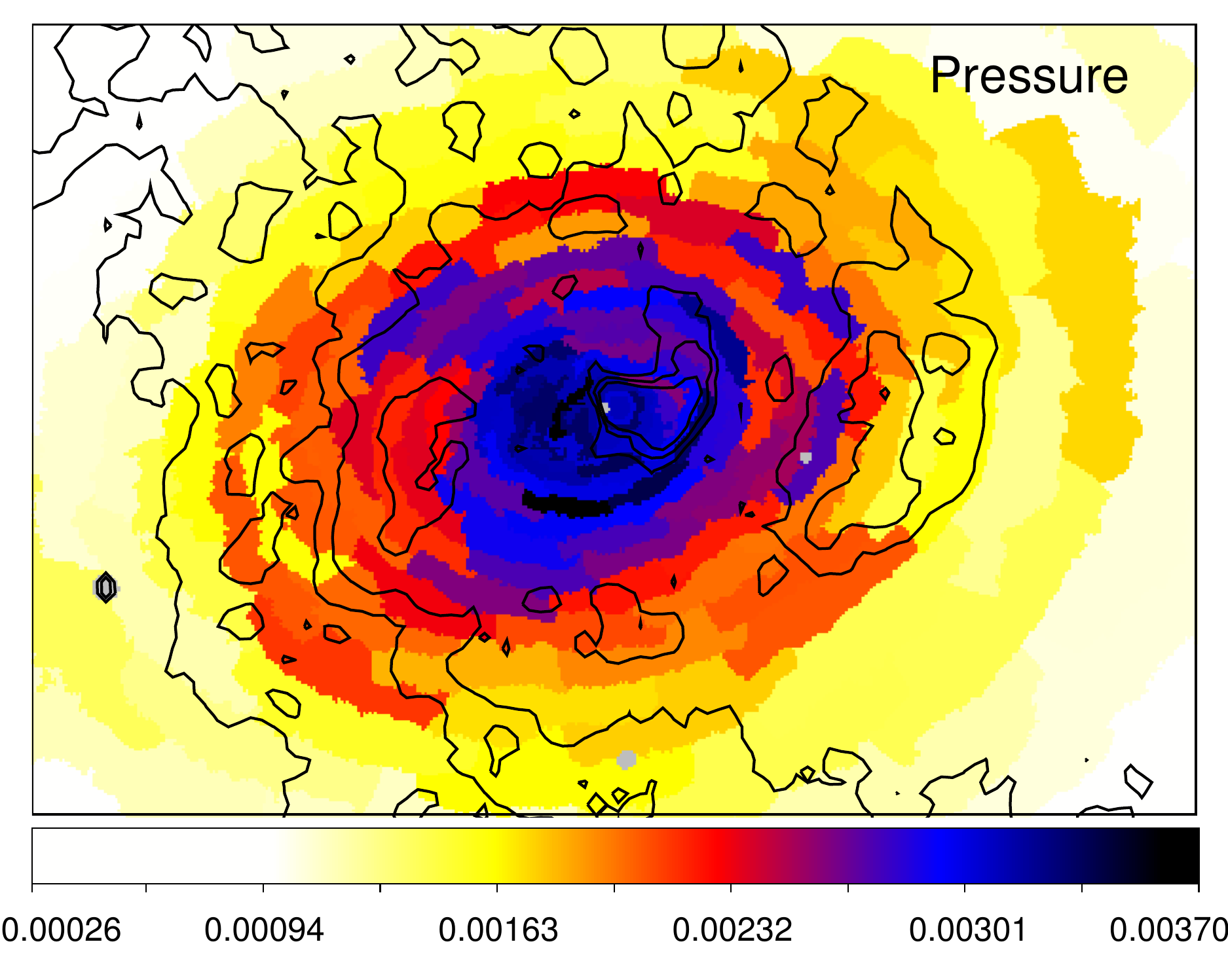}
  \caption{Spectral maps of the inner $206 \times 136$ arcsec region
    ($\sim 390 \times 260$ kpc). (Top panel) Image and contours are the
    fractional differences between an elliptical $\beta$
    model fit to the surface brightness and an adaptively smoothed
    map. (Centre panel) Temperature map of the cluster, using bins
    with a signal to noise ratio of 66. (Bottom panel) Pseudo-pressure
    map of cluster, the square root of the mean surface brightness in
    each bin and multiplied by the temperature.}
  \label{fig:centremaps}
\end{figure}

We now examine the properties of the ICM in the inner $390 \times 260$
kpc. We fitted the spectra from bins with a signal to noise ratio of
66 ($\sim 4350$ counts). In these fits the absorbing column density
and metallicity were allowed to be free parameters. In
Fig.~\ref{fig:centremaps} (top panel) we show the fractional
difference between the $\beta$ model fit to the cluster surface
brightness (see Fig.~\ref{fig:resid}) and an adaptively smoothed image
of the cluster. The image was adaptively smoothed with a top-hat
kernel dynamically adjusted to contain a signal to noise ratio of 30
(900 counts). The contours from this map are also plotted on this and
the other panels. The middle panel shows the best fitting
temperature. The uncertainties in this map range from 10 per cent per
bin in the centre to 30 per cent in the hottest regions.

The surface brightness of the image is proportional to the integral of
the density-squared along the line of sight (with a weak temperature
factor, depending on the X-ray band chosen). If we bin the surface
brightness into our spectral fitting regions, take the square root and
multiply by the temperature, this gives a pseudo-pressure, related to
an average of the pressure with a line of sight factor (bottom panel
of Fig.~\ref{fig:centremaps}). The pressure is asymmetric in the inner
arcmin, with a difference in pressure of $\sim30$ per cent between
the north-west and south-east.

The central temperature map shows some features in common with the
spiral structures seen in the residuals from the $\beta$ model. For
instance there is a cool blob of material to the west of the core
coincident with the brightest region of X-ray emission
(Fig.~\ref{fig:bands}). This region has lower pressure than other
regions at similar radius. At larger radii beyond this western region
is an edge in surface brightness, which we examine in more detail in
Section \ref{sect:edges}. There is another smaller region of cooler
material to the east, coincident with a smaller region of enhanced
X-ray emission and lower thermal pressure.

\section{Radial profiles}
\label{sect:radial}
We investigate the radial profiles of several thermodynamic
quantities. We compare the results of a new method, using multiband
surface brightness projection against more conventional methods.

\subsection{Multiband surface brightness projection}
Our multiband surface brightness projection method (called MultiBand
PROJector or \textsc{mbproj}) builds on the well-known surface
brightness deprojection method of \cite{Fabian81}. It is similar to
the recently published multiband projection model of \cite{Humphrey13}
which analyses surface brightness profiles assuming functional forms
for the mass distribution and entropy profile, inferring the density
and temperature assuming hydrostatic equilibrium. Our technique
differs by using a non-parametric density profile instead of a
parametric entropy profile (although a parametric density profile can
be used) and solves hydrostatic equilibrium by integrating the
pressure inwards to the centre of the cluster.  The aim of the method
is to compute cluster thermodynamic property profiles from count
profiles, without having to do spectral fitting and allowing high
spatial resolution. As the code is forward-fitting, matching a model
to observed data, we call this a projection method, rather than a
deprojection method.

The data used by the procedure are count or surface brightness profiles in
several energy bands. In our data analysis we used three bands. These bands
should be chosen to contain sufficient counts and to be sensitive to
temperature variation. For simplicity in this procedure, we assume
that the surface brightness profiles are extracted from contiguous 2D
annuli on the sky which have the same radii as the 3D shells in the
cluster for which we model the thermodynamic properties. In addition,
we assume constant values for the metallicity and Galactic
absorption. The metallicity requirements can be broken by introducing
parameters into the model representing its profile.

We assume the gravitational potential (excluding the contribution from
X-ray gas mass) has a particular form, the cluster is in hydrostatic
equilibrium and is spherical. To start our analysis, we take this
potential (with initial parameters), an outer log pressure and an
initial log electron density profile, assuming the density is constant
within each shell. A separate density can either be given for each
shell or generated parametrically, for example using a $\beta$
model. We then compute the surface brightness profile in several
energy bands. This procedure is as follows.
\begin{enumerate}
\item The outer pressure and outermost density value is used to
  compute an outermost temperature.
\item The temperature, metallicity and density are converted into an
  emissivity for the outermost shell in the cluster in several energy
  bands. This calculation is done using \textsc{xspec} with the
  \textsc{apec} spectral model, given a response matrix and ancillary
  response matrix. For purposes of efficiency, the temperature to
  emissivity conversion is precalculated using a grid of temperature
  values for each band for unit emission measure. The calculation is
  done with zero and Solar metallicity, using interpolation to
  calculate the results using other metallicity values. The effects of
  Galactic photoelectric absorption are included in the emissivities,
  although this prevents modelling variation across the source.
\item \label{enum:loop} We now consider the next innermost shell. We
  compute a pressure for this shell from the sum of the previous
  pressure plus the contribution from the hydrostatic weight of the
  atmosphere, i.e. $\delta P = \delta r \: \rho \: g$, where $\delta
  r$ is the radial spacing and $\rho$ is the mass density. $g$ is the
  gravitational acceleration computed from the potential in the shell
  plus a contribution to the acceleration from the gas mass of
  interior shells. $g$ for the potential is calculated at the
  gas-mass-weighted-average radius for the shell, assuming constant
  gas density
\begin{equation}
  r_\mathrm{weighted} = \frac{3}{4} \frac{
  (r_\mathrm{in}+r_\mathrm{out})(r_\mathrm{in}^2+r_\mathrm{out}^2) }
  {r_\mathrm{in}^2 + r_\mathrm{in} r_\mathrm{out} + r_\mathrm{out}^2},
\end{equation}
where $r_\mathrm{in}$ and $r_\mathrm{out}$ are the interior and
exterior radii of the shell. To calculate the gas mass contribution to
$g$, we calculate the total gravitational force on the shell from the
mass of interior shells ($M_\mathrm{int}$) and from the shell itself,
at constant density $\rho$. This is then divided by the total mass
within the shell to compute an average acceleration,
\begin{equation}
  g_\mathrm{gas} = G \frac { 3 M_\mathrm{int} + \rho(r_\mathrm{out} -
    r_\mathrm{in})[(r_\mathrm{out} + r_\mathrm{in})^2 + 2r_\mathrm{in}^2] } 
  {r_\mathrm{in}^2 + r_\mathrm{in} r_\mathrm{out} + r_\mathrm{out}^2}.
\end{equation}
\item Taking this pressure and the density value in this shell, we
  compute its emissivity in the several bands.
\item We go back to step \ref{enum:loop} until we reach the innermost
  shell.
\item Given the emissivities for each shell, we compute the projected
  surface brightness in each band by multiplying by the volumes of
  each shell projected onto annuli on the sky and summing along the
  line of sight. We include the sky background by adding a constant
  onto each surface brightness profile.
\end{enumerate}

Once we have the surface brightness profiles in each band for our
choice of parameters (parameters of the gravitational potential, outer
pressure and temperature profile) we can compare them to the observed
surface brightness profiles. The projected surface brightness profiles
are multiplied by the areas on the sky of each annulus and the
exposure time of the observation to create count profiles. The
exposure times can also include an additional factor from the average
difference in an exposure map between the annulus and where the
response matrices are defined, to account for detector features and
vignetting. If sectors are examined within the cluster, rather than
full annuli, this can be included within the area calculation, along
with the negative contribution of any excluded point sources. To
assess how well the model fits the data, we calculate the logarithm of
the likelihood of the model for Poisson statistics \citep{Cash79} with
\begin{equation}
  \log \mathcal{L} = \sum_{i}{ \left[ d_i \log m_i  - m_i - \log \Gamma(d_i+1) \right]},
\end{equation}
where $d_i$ and $m_i$ are the observed and predicted number of counts,
respectively. $i$ indexes over each annulus in each band.

To obtain the initial density profile, we add the surface brightness
profiles from all the energy bands and deproject the surface
brightness profile assuming a particular temperature value (here 4
keV) and spherical symmetry. The initial outer pressure is the outer
density times the temperature value. We then fit the complete set of
parameters, including the potential parameters, by maximising $\log
\mathcal{L}$ using a least-squares minimisation routine. The
basin-hopping algorithm \citep{Wales97} is then employed to help
ensure that the fit is not in a local minimum.

We start an MCMC analysis to obtain a chain of parameter values after
a burn-in period. We again use \textsc{emcee} for the MCMC
analysis. The initial parameters for the walkers are set to be tightly
clustered around the best fitting parameters. Flat priors are used on
all parameters.  After the run, uncertainties on the parameters can be
obtained by computing the marginalised posterior probabilities from
the chain for each parameter. To compute radial profiles of other
thermodynamic quantities, we take sets of parameters from the chain
and compute the quantity for each set of parameters and examine the
distribution of these output values. We compute the electron density,
electron pressure, electron entropy (defined as $K_\mathrm{e} =
n_\mathrm{e} T^{-2/3}$), cumulative gas mass, cumulative luminosity,
mean radiative cooling time, cumulative mass deposition rate in the
absence of heating (accounting for the gravitational contribution;
\citealt{Fabian94}), gravitational acceleration, total cumulative mass
and gas mass fraction.

\subsubsection{Rebinning input profiles}
The sizes of the input annular bins must be chosen in some way. When
fitting for separate densities in each shell, there is one density
parameter per shell. To reduce the uncertainties on the densities, the
input profiles need to be binned. In principle, if a functional form
is assumed for the density profile then the input profiles do not need
binning, but in practice to get a good initial density estimate, some
binning is required.

A second code, \textsc{mbautorebin}, is used to rebin the profiles
before analysis. We take initial count profiles for each band,
computed using pixels. The profiles are added to give a total count
profile. The method works inwards from the outside of the profile.
Radial annular bins are combined until the fractional uncertainty on
the emissivity in the respective 3D shell drops below a threshold and
the total number of counts in the shell is greater than a second
threshold. Once the threshold is reached, we then consider the next
innermost bin.

The emissivity in a shell is computed by deprojecting the counts in
each annulus assuming spherical symmetry. To be precise, to calculate
the emissivity, the projected contributions from the background and
already binned 3D shells is subtracted from the count rate for the
annulus, and this rate is then divided by the volume of the shell
within the annulus. Monte Carlo realisations of the count rate in the
shell and those shells outside are used to estimate the uncertainty.

For the central remaining bin after this process, we combine it with
the next innermost bin if the uncertainty on the emissivity is
substantially below the threshold.

There may be some form of bias generated from this procedure if the
number of counts is very low. Bin sizes may depend on the radii of
particular counts. In the low count regime (a few hundred counts or
less), we used an alternative version of this code which fits a
$\beta$ model to the surface brightness profile. The bin radii are
adjusted to give the same fractional uncertainty on the deprojected
emissivity in each bin, based on realisations of the $\beta$ model.

\subsubsection{Null potential case}
The above procedure can be modified to compute thermodynamic
profiles without the assumption of hydrostatic equilibrium. We
  term this adaptation of the model the Null potential case. Here we
add parameters for the temperature in each shell and do not use the
gravitational potential. Rather than calculating the pressure assuming
hydrostatic equilibrium, we multiply the temperature and density
values. This variation of the model is in essence a low-spectral
resolution version of spectral fitting projection models, such as
\textsc{projct}. It provides a useful check to ensure that the
thermodynamic profiles are not being biased by the choice of
gravitational potential or by parts of the cluster being out
  of hydrostatic equilibrium.

\subsection{Complete radial profiles}
\label{sect:completeprof}
To examine the average cluster profiles, we extracted count profiles
for the cluster from images in the 0.5 to 1.2, 1.2 to 2.5 and 2.5 to 6
keV bands out to a radius of 4.55 arcmin (516 kpc). The profiles were
extracted on a pixel-by-pixel basis and the areas calculated by
summing numbers of pixel. A 0.5 to 7 keV exposure map was used to
correct the areas for bad pixels and other detector features. We
restricted the analysis to region covered by the ACIS-S3 CCD for the
12881 observation, excluding point sources. In addition to using
standard background event files to obtain the X-ray and particle
backgrounds, we account for out-of-time events which occur during
detector readout using the \textsc{make\_readout\_bg} script of
M. Markevitch to generate out-of-time event files. Out-of-time events
would otherwise lead to regions along the readout direction being
contaminated with emission from the centre of the cluster. PKS0745 has
a bright central core, making this problem more obvious. We extracted
surface brightness profiles from both background event files,
combining them. The foreground and background input profiles were
rebinned to obtain a maximum deprojected emissivity error of 4 per
cent. This gives around 20\,000 projected counts per radial bin over
most of the radial range, reducing to around 10\,000 near the
outskirts.

\begin{table*}
  \caption{Details of the MCMC analysis of the various data and samples.
    Shown are the data analysed, the potential used, the maximum log
    likelihood, number of walkers used, burn length used, sample length
    used, acceptance fraction, mean autocorrelation period
    ($\langle\tau\rangle$) and maximum autocorrelation period ($\max \tau$).}
  \begin{tabular}{cccccccccc}
    \hline
    Region & Potential & $\max \log \mathcal{L}$ & Walkers & Burn & Length & Acceptance & $\langle\tau\rangle$ & $\max \tau$ \\ \hline
    Full   & NFW       & $-941.7$                & 200     & 1000 & 4000   & 0.18       & 134                  & 311         \\
           & King      & $-865.8$                & 200     & 1000 & 4000   & 0.18       & 131                  & 302         \\
           & Arb4      & $-861.7$                & 400     & 2000 & 8000   & 0.18       & 214                  & 591         \\
           & Null (bin 2) & $-846.2$             & 400     & 2000 & 8000   & 0.15       & 259                  & 611         \\
           & Null      & $-833.9$                & 800     & 8000 & 90000  & 0.08       & 2170                 & 5293        \\ 
           & NFW $>0.5'$ & $-638.8$              & 200     & 1000 & 4000   & 0.20       & 118                  & 282         \\
           & King $>0.5'$ & $-636.7$             & 200     & 1000 & 4000   & 0.21       & 116                  & 295         \\ \hline
    West   & NFW       & $-451.4$                & 200     & 1000 & 4000   & 0.22       & 110                  & 294         \\
           & King      & $-448.8$                & 200     & 1000 & 4000   & 0.22       & 102                  & 277         \\ \hline
    East   & NFW       & $-445.7$                & 200     & 1000 & 4000   & 0.21       & 113                  & 297         \\
           & King      & $-445.0$                & 200     & 1000 & 4000   & 0.21       & 109                  & 291         \\ \hline
  \end{tabular}
  \label{tab:mcmc}
\end{table*}

\begin{figure*}
  \includegraphics[width=\textwidth]{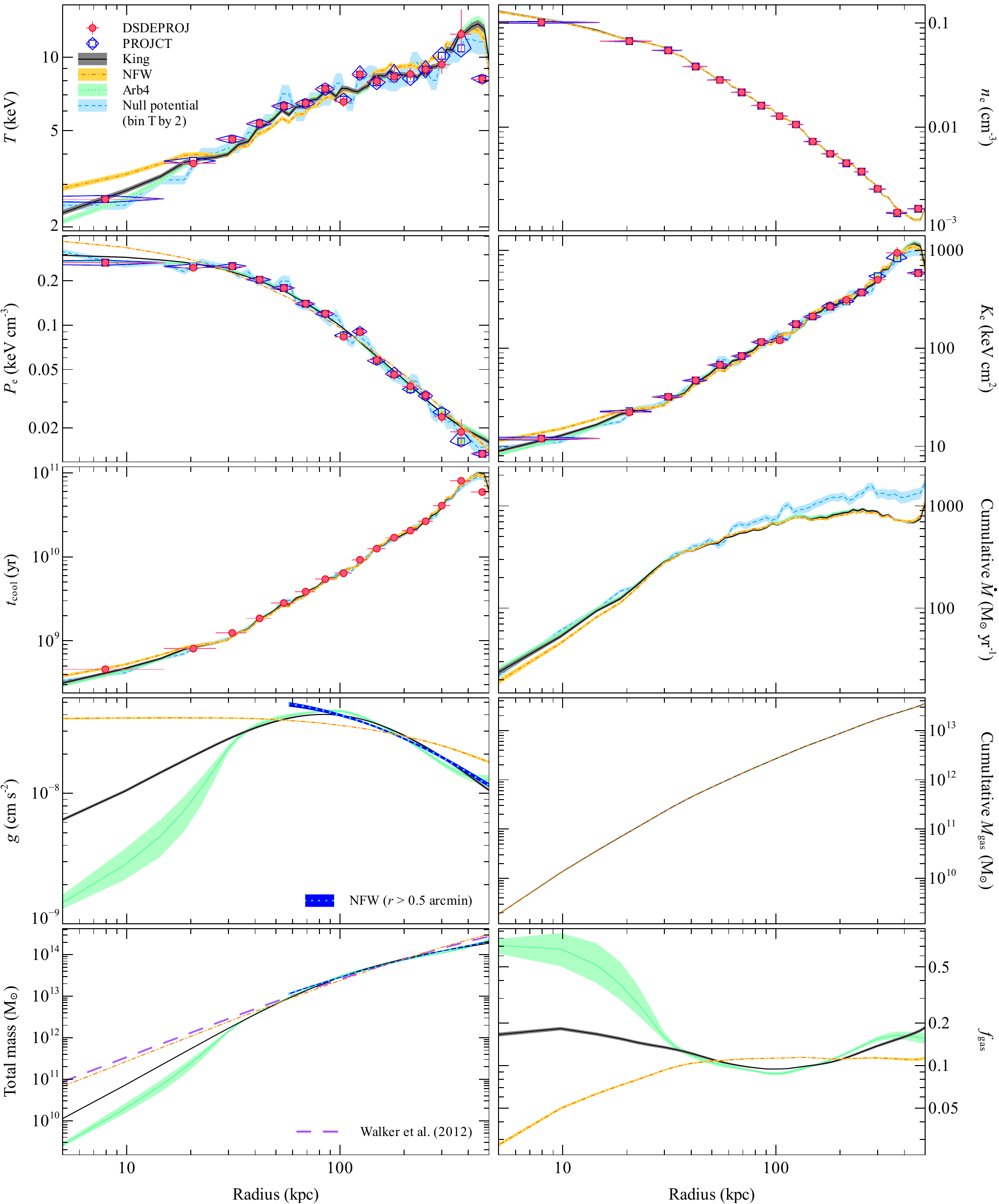}
  \caption{Complete radial profiles, comparing analysis
    methods. Shown are the results for spectral fitting using
    \textsc{dsdeproj} and \textsc{projct}, and \textsc{mbproj} with
    several mass models, including a King, an NFW, a 4 bin
    parametrisation (Arb4) and a null potential (fitting densities
    instead). The quantities shown include the temperature ($T$),
    electron density ($n_\mathrm{e}$), electron pressure
    ($P_\mathrm{e}$), electron entropy ($K_\mathrm{e}$), mean
    radiative cooling time ($t_\mathrm{cool}$), cumulative mass
    deposition rate ($\dot{M}$), gravitational acceleration ($g$) and
    cumulative gas mass ($M_\mathrm{gas}$). In the $g$ and total mass
    plots we also show the results for an NFW model which only
    examines the data outside 0.5 arcmin radius. For the total mass
    panel, we show the mass profile obtained from \emph{Suzaku} data
    by \protect\cite{Walker12}.}
  \label{fig:deprojprofs}
\end{figure*}

Several different potential models were examined using
\textsc{mbproj}, with the results shown in
Fig.~\ref{fig:deprojprofs}. The central lines show the median parameter
values taken from the chain, with shaded bar showing the uncertainties
taken from the 15.9 and 84.1 percentiles, equivalent to $1\sigma$
errors. The potentials used were an NFW model \citep{NFW96}, a King
model \citep{King62}, the null potential and an arbitrary potential,
Arb4, where we parametrised the mass density at 4 logarithmically
spaced radii over the radial range and used linear interpolation and
integration to compute the gravitational acceleration. We assumed a
metallicity of $0.4 \Zsun$ and an absorbing column density of $3.78
\times 10^{21} \psqcm$, based on an average of values from fitting
bins over the cluster core. For the null potential model we show the
results where there is a single temperature parameter in each shell
and one where the temperatures in each pair of bins is assumed to be
the same (bin 2).

The details of the MCMC analysis are shown in Table \ref{tab:mcmc}. We
show log likelihoods for the best fitting parameters, the number of
walkers used, the burn period, the length of the final chain. The
acceptance fraction of the chain is shown. Ideally this should be
between 0.2 and 0.5 \citep{ForemanMackey12}. Low values, as seen in
the null potential model with a single temperature parameter for each
bin, may indicate multi-modal parameter values which are difficult for
the MCMC analysis to sample.

\begin{table}
  \centering
  \caption{Derived marginalised  posterior
    probabilities for the parameters on the potentials from the surface brightness analysis.
    For the Arb4 model the density units are $\log_{10} 10^{-24} \gpcm$.}
  \begin{tabular}{lll}
    \hline
    Potential & \multicolumn{2}{c}{Parameters} \\ \hline
    NFW       & $c=3.88 \pm 0.14$             & $r_{200} = (2.20 \pm 0.06) \Mpc$ \\
    King      & $\sigma = (769 \pm 9) \kmps$  & $r_\mathrm{core} = (83 \pm 2) \kpc$ \\
    Arb4      & $\rho_{1}=-1.0^{+0.3}_{-0.6}$ & $\rho_{2} = 0.05^{+0.05}_{-0.04}$ \\
              & $\rho_{3}=-0.45 \pm 0.03$     & $\rho_{4} = -1.80 \pm 0.08$ \\
    NFW ($r>0.5'$) & $c = 9.2_{-0.8}^{+0.5}$  & $r_{200} = 1.42_{-0.03}^{+0.05} \Mpc$ \\
    King ($r>0.5'$)& $\sigma = (760 \pm 12) \kmps$ & $r_\mathrm{core} = (78 \pm 6) \kpc$ \\
    \hline
  \end{tabular}
  \label{tab:massparams}
\end{table}

For models with more parameters (the Null and Arb4 models), we
increased the number of walkers in the analysis. Table \ref{tab:mcmc}
also shows the mean and maximum autocorrelation periods ($\tau$) for
parameters from individual walkers in the chain. To properly sample
parameter values so that the chain has converged, the chain should be
$\sim 10$ times longer than this period \citep{ForemanMackey12}. This
is the case on our analyses. The null models have long autocorrelation
periods, likely due to the multivalued temperatures at larger radii,
and so required much longer chain lengths. The derived marginalised
posterior probabilities for the mass models are given in Table
\ref{tab:massparams}.

\begin{figure}
  \includegraphics[width=\columnwidth]{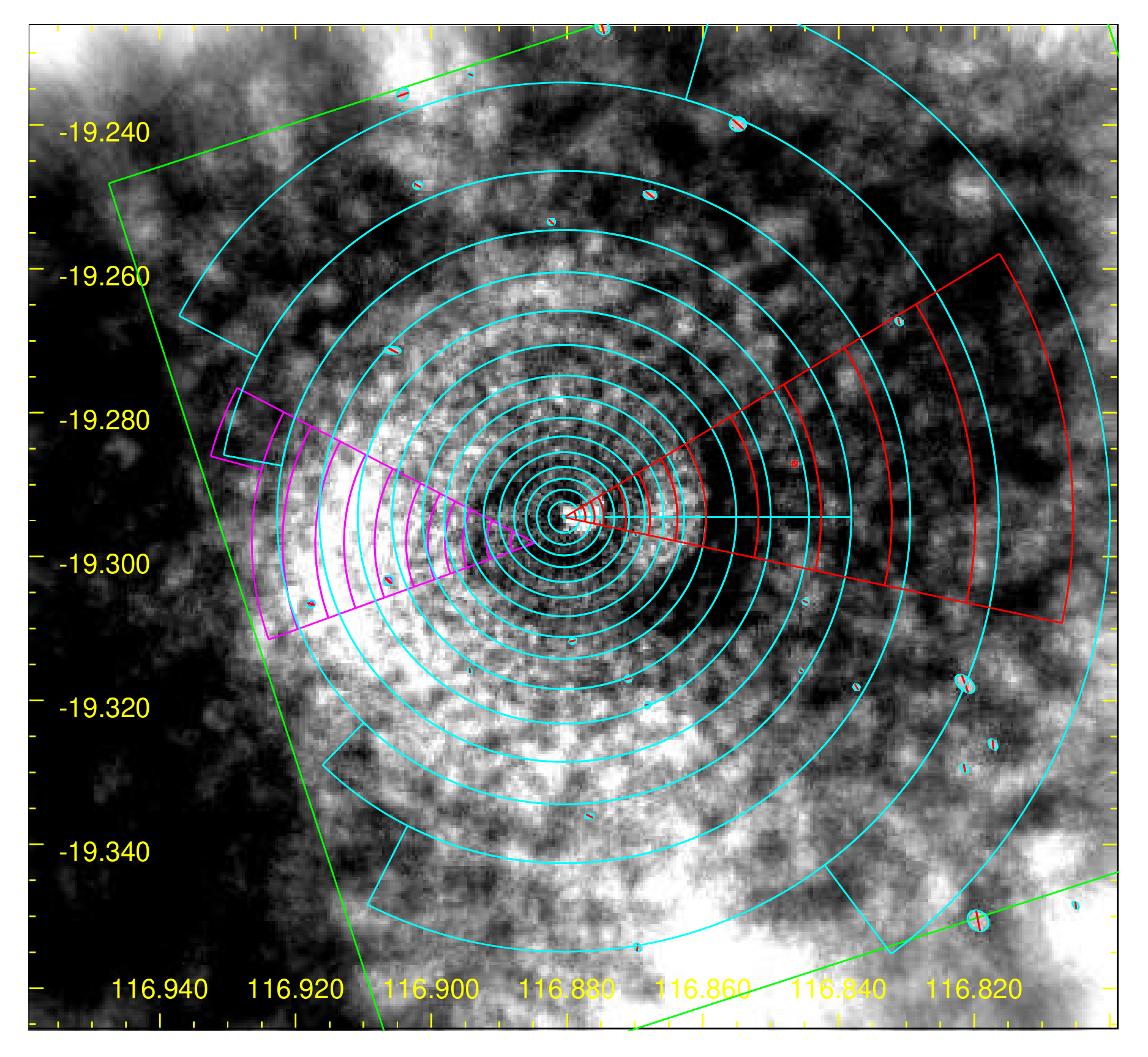}
  \caption{$\beta$-model-residual image showing the
    annular sectors used for the \textsc{dsdeproj} and \textsc{projct}
    spectral analyses of the whole azimuthal range and the western and
    eastern sectors. The rotated box shows the edge of the ACIS-S3 CCD
    in the 12881 observation. The X-ray image was adaptively
    smoothed to have a signal to noise ratio of 15 in the smoothing
    kernel and was then divided by an elliptical $\beta$ model fit.}
  \label{fig:sectors}
\end{figure}

To compare the results from our analysis, we also examined the results
of spectral fitting using two different methods for account for
projection effects. The first method was to use the \textsc{dsdeproj}
code to deproject projected spectra \citep{SandersPer07,Russell08}. We
also used the \textsc{projct} model in \textsc{xspec}, which projects
spectra.  The radii of the annuli chosen were the surface brightness
deprojection bins, but binned up by a factor of three
(Fig.~\ref{fig:sectors}). We note that for these spectral fits, we
fixed the column density to the same value as the surface brightness
deprojection. The metallicity of each shell, however, was free in the
spectral fits. The spectra were fitted between 0.5 and 7 keV. For
\textsc{projct}, all the spectra were fitted simultaneously to
minimise the total $\chi^2$ (the reduced $\chi^2$ was
$1.02=6447/6321$).

The spectral results are compared with the data in
Fig.~\ref{fig:deprojprofs}. There appears to be good agreement between
the spectral fitting results and our deprojection code. However, there
is a discrepancy in the temperature and pressure results at small and
large radius between the NFW analysis and the other results. Examining the
gravitational acceleration, $g$, the Arb4 gravitational model appears
to match the King model fairly well except at small radii, but matches
the NFW model poorly.  If we exclude the central region from the data
and only fit the data beyond 0.5 arcmin radius, the NFW model matches
the other models much better in that region (bottom left two panels of
Fig.~\ref{fig:deprojprofs}).

Similar results to the Arb4 model are also found with a model where
the gravitational acceleration is parametrised in 6 logarithmic radial
intervals, using linear interpolation to calculate intermediate
values. In this case, the gravitational acceleration is unconstrained
inside 10~kpc radius, but $g$ is similar to the Arb4 model elsewhere.

We also note that the null potential model shows different mass
deposition rates ($\dot{M}$) compared to the other models because it
does not include the gravitational contribution. Both the spectral and
projection methods have anomalous values in their outer bins. This is
due to cluster emission outside these radii not being taken into
account.

\subsection{Surface brightness edges}
\label{sect:edges}

In the western half of the cluster, there are two edges visible in the
X-ray surface brightness (Fig.~\ref{fig:resid}). We have analysed the
thermodynamic properties across these edges by spectral fitting and by
applying \textsc{mbproj}. The sectors we used for the spectral fitting
are shown in Fig.~\ref{fig:sectors}. The same azimuthal and radial
ranges were used for the surface brightness analysis. The edges of the
spectral extraction regions were chosen to match the edges in surface
brightness as closely as possible. We analysed the data similarly to
Section \ref{sect:completeprof} but using a column density of $4.27
\times 10^{21} \pcmsq$. This increased value was necessary due to the
westward rise in column density (Section \ref{sect:largemaps}) and was
obtained as the best fitting value from the \textsc{projct}
analysis. The surface brightness profiles were rebinned in radius to
have a minimum uncertainty on the emissivity of 6 per cent. We used
both King and NFW mass models in our \textsc{mbproj} analysis.

\begin{figure}
  \includegraphics[width=\columnwidth]{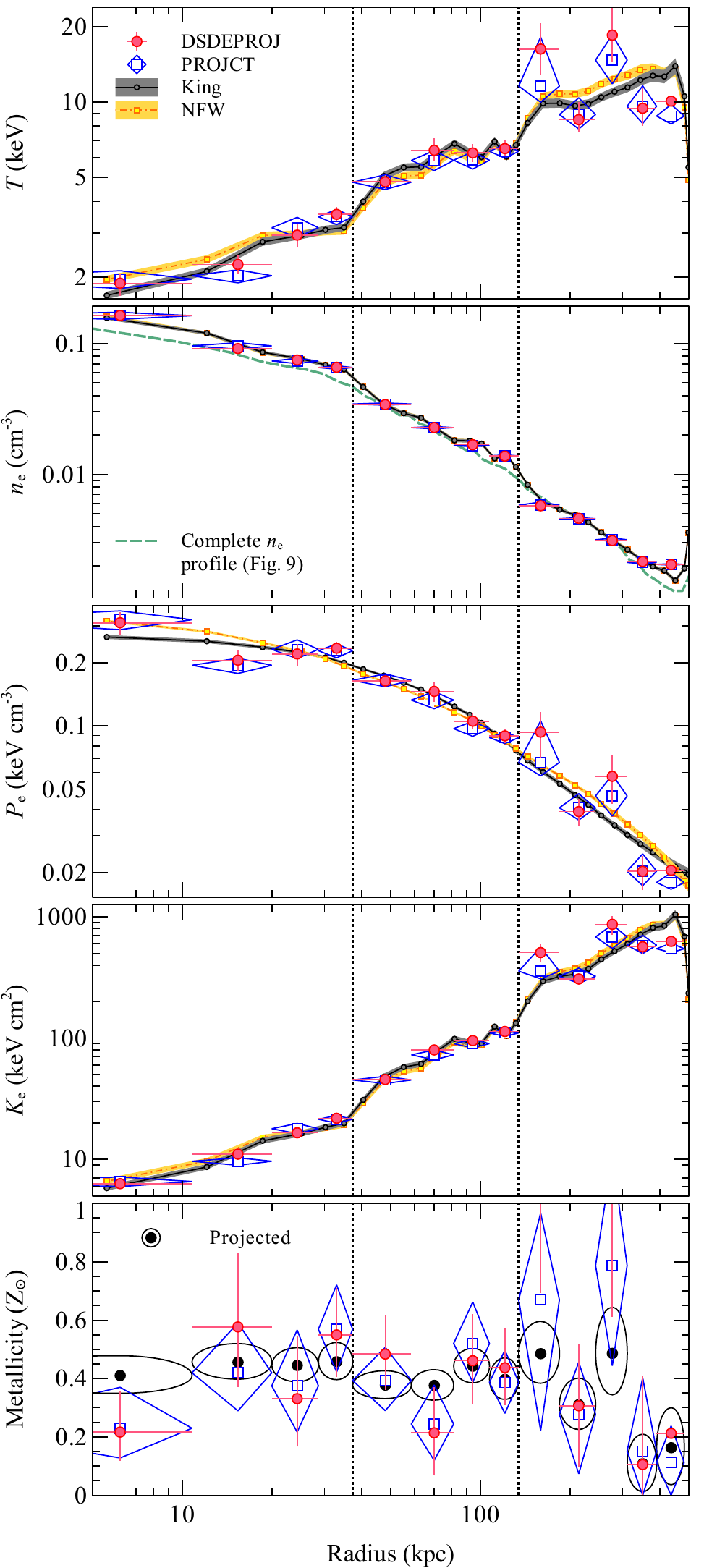}
  \caption{Profiles to the west of the core across the two surface
    brightness edges, of temperature, electron density, electron
    pressure, electron entropy and metallicity. The positions of the
    edges are marked by dotted lines.  Shown are the results for
    spectral fitting (\textsc{dsdeproj}, \textsc{projct} and Projected
    for metallicity) and using \textsc{mbproj} (with NFW and King
    potentials). The complete cluster results from
      Fig.~\ref{fig:deprojprofs} are also plotted with the electron
      density profile.}
  \label{fig:jumpprofile}
\end{figure}

The profiles in this sector are shown in
Fig.~\ref{fig:jumpprofile}. The results show that there are two
obvious breaks in density, temperature and entropy at the locations of
these edges. The \textsc{mbproj} results suggest that the breaks have
finite width, although any perturbations in the shape of the edge
relative to our extraction region will broaden its measured width. The
spectral fitting results indicate that there may be a break in
pressure at the innermost edge. Using the \textsc{mbproj} method,
there can never be a discontinuity in pressure as a hydrostatic
atmosphere is assumed. The spectral fitting pressure inside the edge
are enhanced with respect to the smooth curves. Considering only the
two points either side of the edge, the pressure jump relative to the
smooth curves is around $2 \sigma$ in significance. However, if
powerlaw models were fitted inside and outside the edge, then the
significance of the break would increase.  The metallicity profile in
the centre is remarkably flat at $0.4\Zsun$ within 300~kpc radius.

\begin{figure}
  \includegraphics[width=\columnwidth]{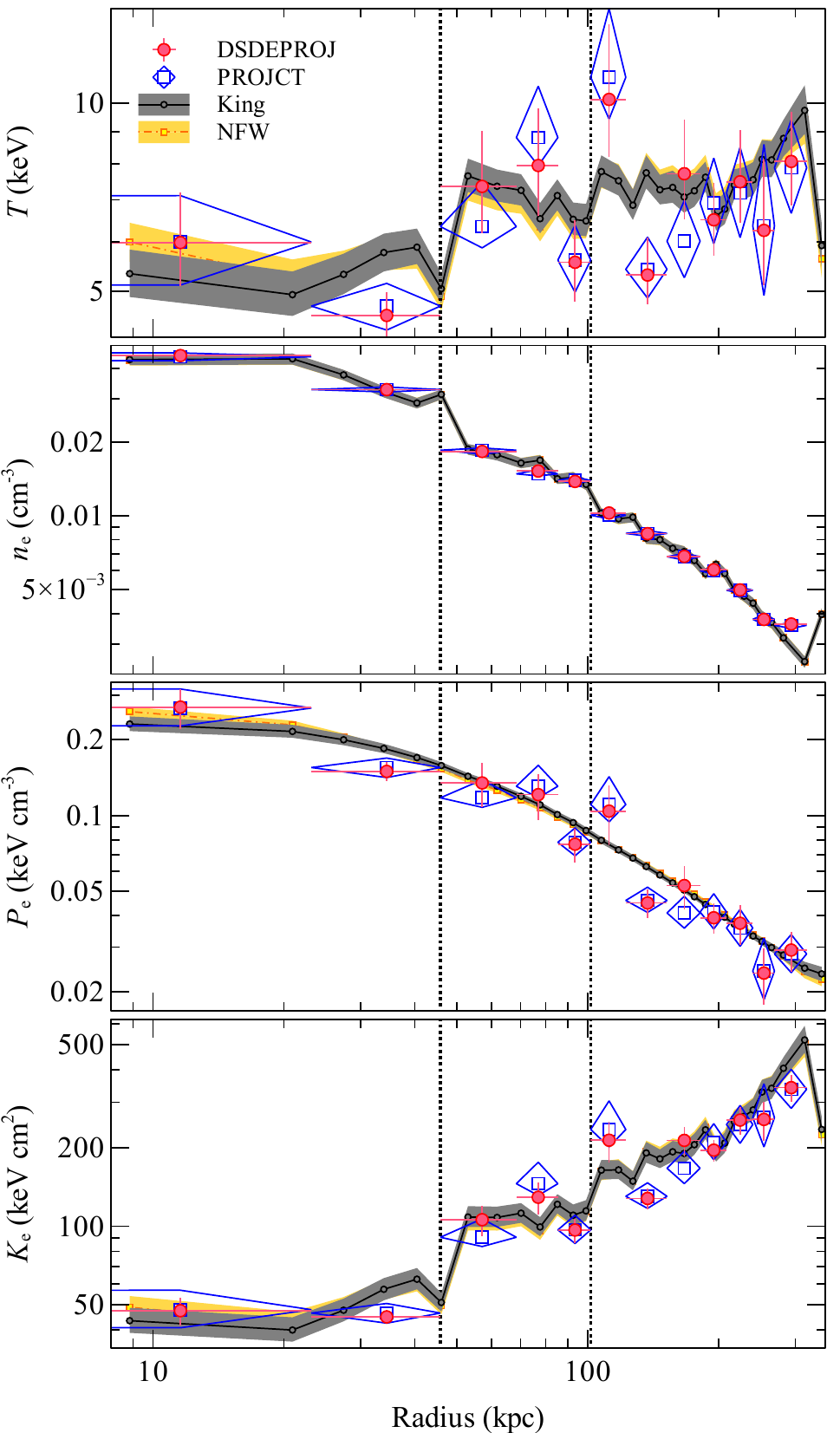}
  \caption{Profiles of quantities to the east of the cluster core,
    across two surface brightness edges, marked by vertical
    lines. Note that the centre of this deprojection is offset 20
    arcsec south-east of the core (38 kpc), to better match the sector
    to the edges. We show results for spectral fitting
    (\textsc{dsdeproj} and \textsc{projct}) and \textsc{mbproj} using
    NFW and King potentials.}
  \label{fig:jumpprofilee}
\end{figure}

We similarly examined the properties of the cluster in the eastern
direction across the surface brightness edges
(Fig.~\ref{fig:jumpprofilee}). This is more difficult because the
edges are not smooth or centred on the cluster nucleus. They appear to
look rather kinked (Fig.~\ref{fig:resid}), similar to those seen in
seen in Abell 496 \citep{Roediger12}. We have therefore examined the
deprojected profiles using a sector with a centre offset around 40 kpc
south west from the core of the cluster (Fig.~\ref{fig:sectors}), so
that the circular sectors better match the surface brightness
features. This introduces systematic geometric error into the
analysis. The radial features, however, are also seen in projected
profiles. We repeated the same spectral and \textsc{mbproj} analyses
as in the western sector, but rebinning the input profiles to give a
minimum uncertainty of 10 per cent on the emissivity.

We again see density, temperature and entropy jumps across the two
edges (although the outer temperature jump is less significant). The
pressure does not appear to jump at the edges. Fig.~\ref{fig:sectors}
shows a further reduction with surface brightness near the edge of our
sector (at radii of $\sim 240$ kpc from the cluster core) associated
with the spiral structure. Unfortunately this edge lies close to the
boundary between two neighbouring CCDs, making it difficult to examine
both in terms of surface brightness and spectrally.

\section{Larger scale maps}
\label{sect:largemaps}
\begin{figure*}
  \includegraphics[width=0.48\textwidth]{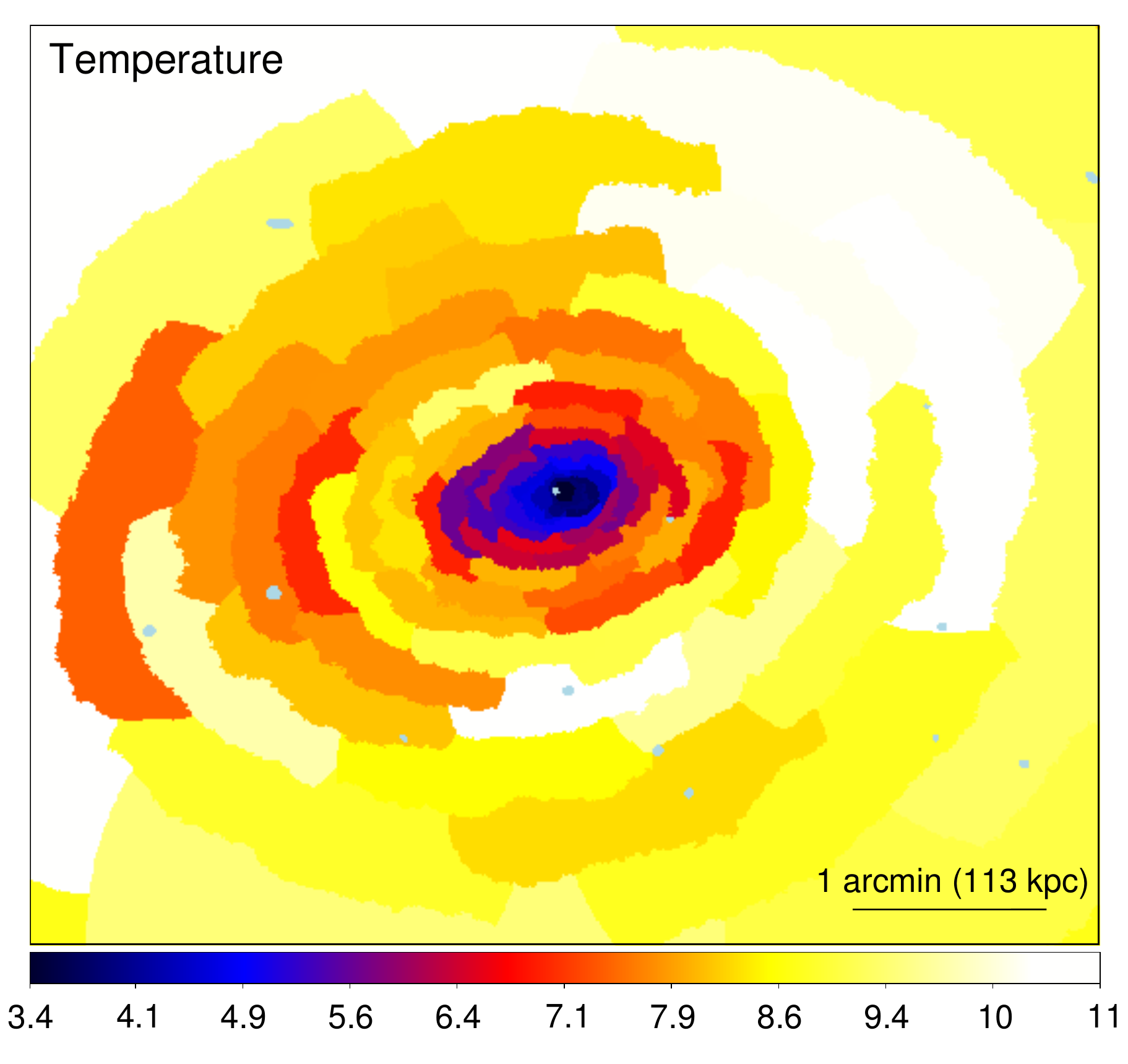}
  \includegraphics[width=0.48\textwidth]{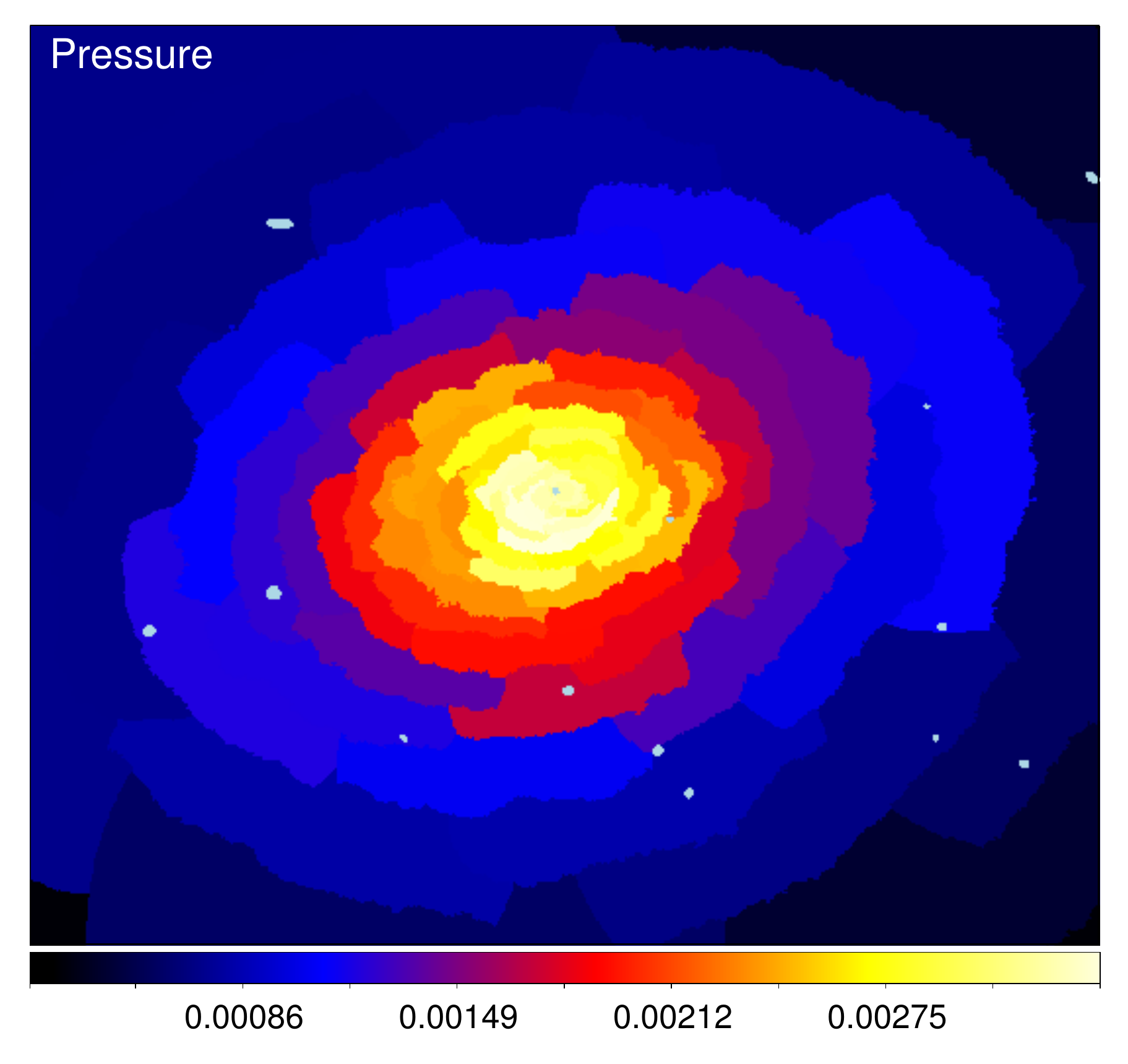}
  \includegraphics[width=0.48\textwidth]{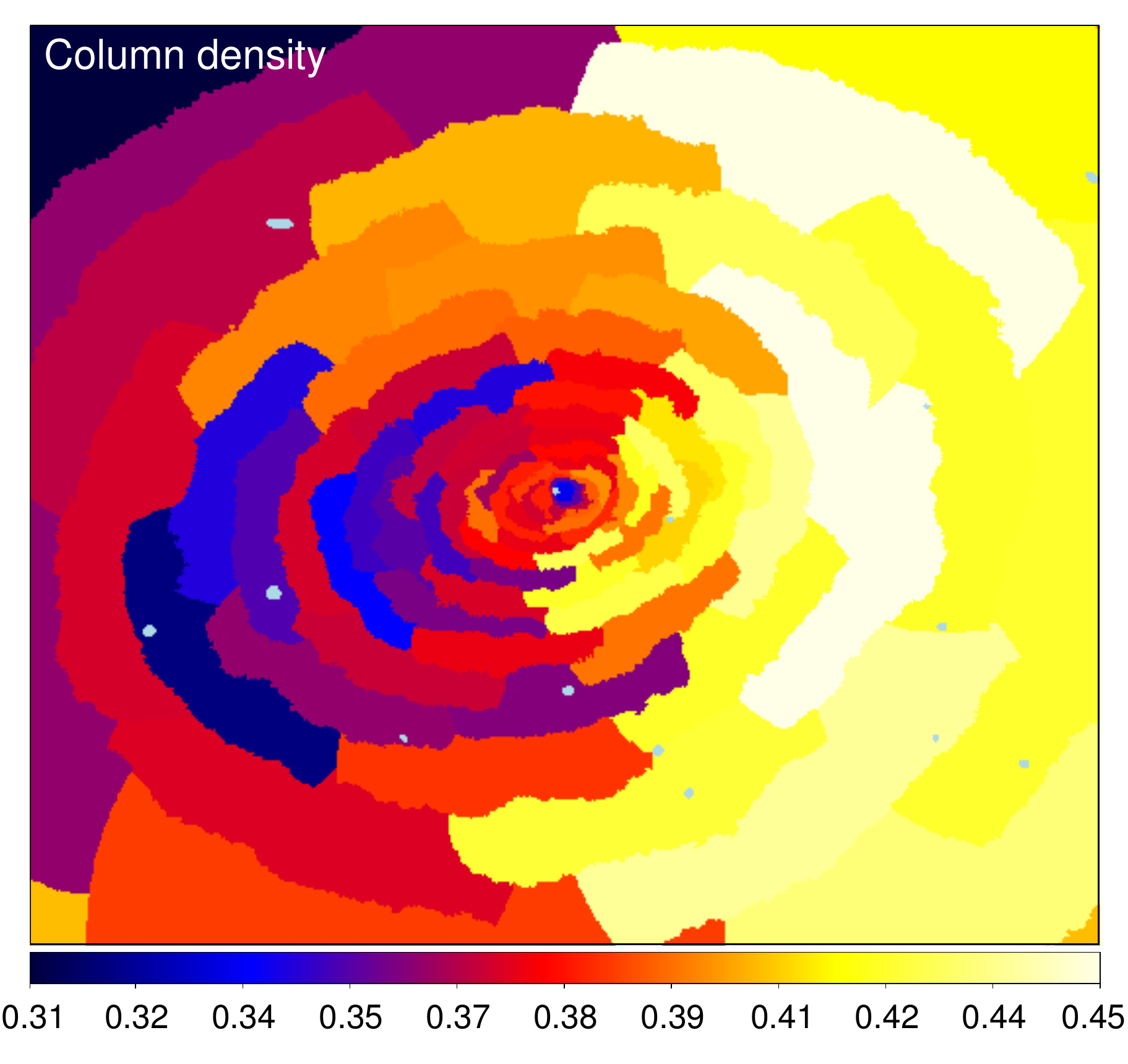}
  \includegraphics[width=0.48\textwidth]{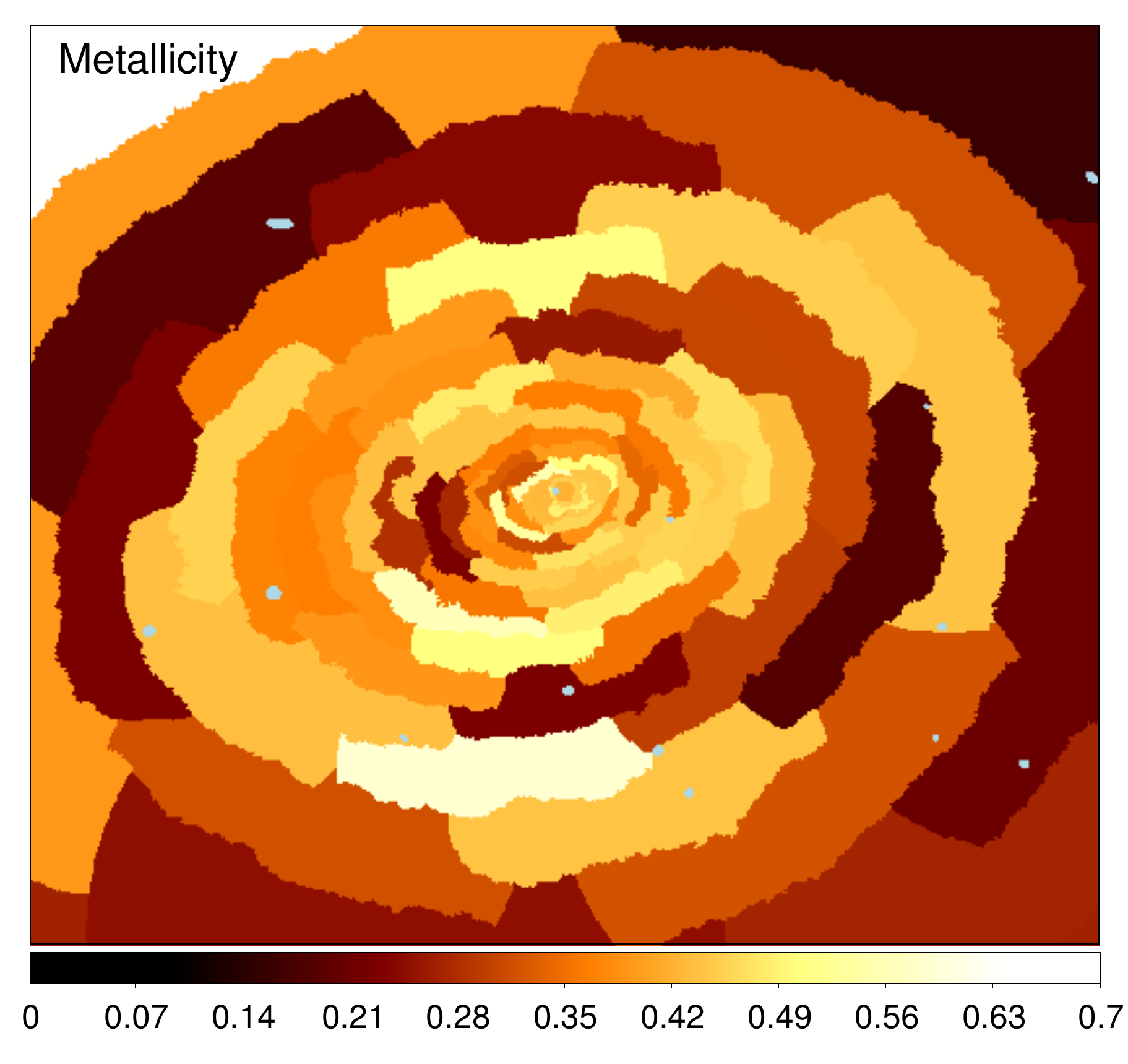}
  \caption{Maps of the temperature in keV (top-left), pseudo-pressure
    (top-right), absorbing column density in $10^{22} \pcmsq$
    (bottom-left) and metallicity in Solar units (bottom-right) on
    large scales. The cluster was binned to have regions with a signal
    to noise ratio of 101 ($\sim 10^5$ counts).}
  \label{fig:largermaps}
\end{figure*}

Fig.~\ref{fig:largermaps} shows larger scale maps of the properties of
the ICM in the cluster. To produce these maps we fitted spectra from
bins with a signal to noise ratio of 101 (10200 counts). We use a
single component APEC model, allowing the temperature, metallicity,
normalisation and absorbing column density to be free in the fits. In
addition to extracting background spectra from
standard-background-event files, spectra were extracted from the
out-of-time-event files for each bin, combining the datasets. In
\textsc{xspec} they were loaded as a correction file during spectral
fitting for each bin. Input spectra were grouped to have a minimum of
20 counts per spectral bin. We fit the spectra between 0.5 and 7 keV,
minimising the $\chi^2$ of the fit.  The resulting maps show that
there is considerable variation in absorbing column density across the
image. The variation is in the direction of galactic latitude, with
stronger absorption towards the Galactic plane. As PKS0745 is near the
galactic plane, this suggests that the variation could be real, as
implied by the large standard deviation of the nearby LAB
values. Alternatively, the variation could be due to calibration, for
example uncorrected contaminant on the ACIS detector. Analysis of
\emph{XMM} data would confirm whether it is real. There are also
non-radial variations in temperature and pressure, which we examine
further in the following section.

\begin{figure}
  \includegraphics[width=\columnwidth]{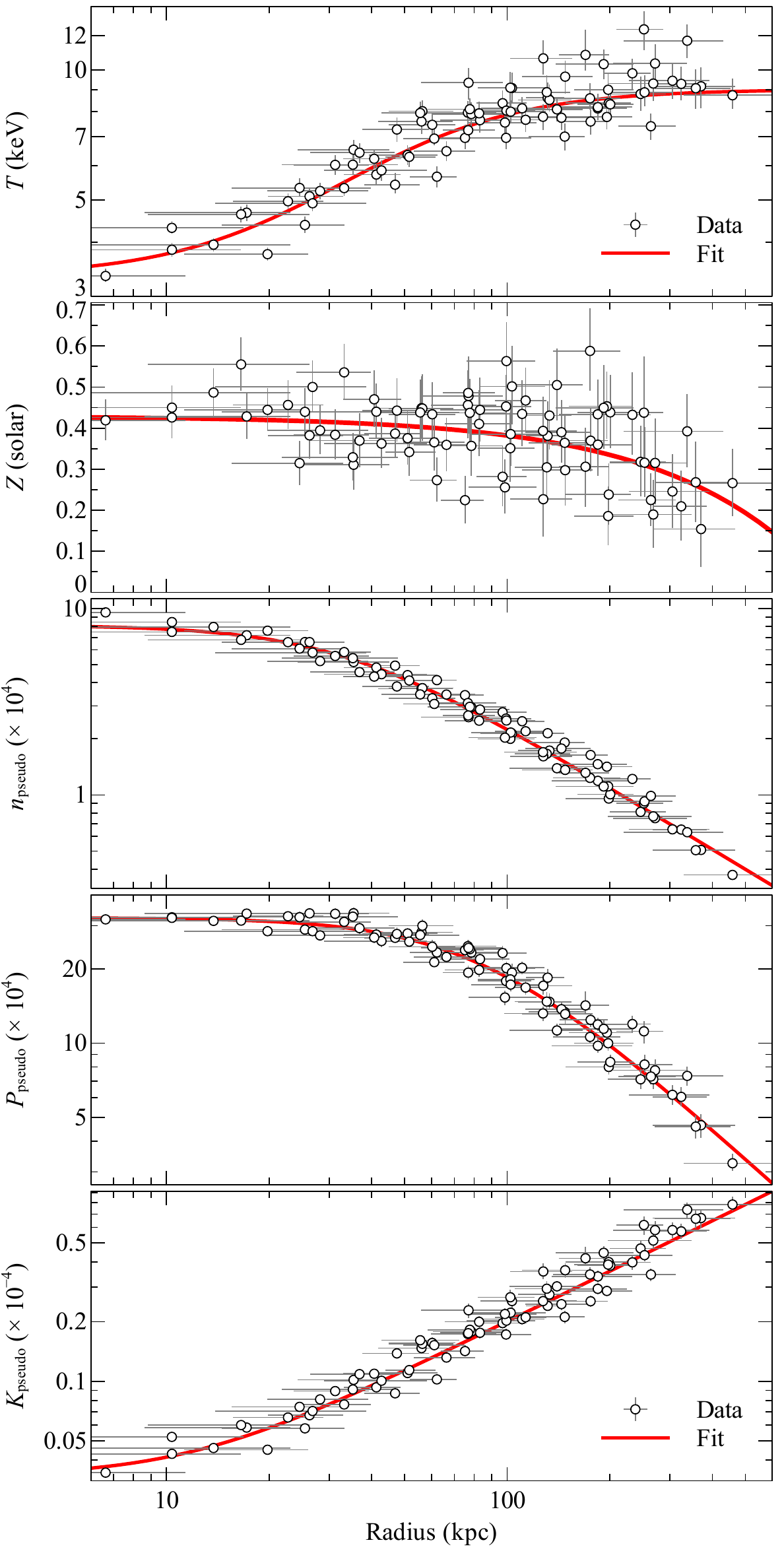}
  \caption{Profiles generated from the maps in
    Fig.~\ref{fig:largermaps}. The value of the bin is plotted against
    the mean bin position. The radial error bars show the bin
    extents. The quantities are temperature (1st panel), metallicity
    (2nd panel), pseudo-density (3rd panel), pseudo-pressure (4th
    panel) and pseudo-entropy (5th panel).}
  \label{fig:map_profiles}
\end{figure}

\begin{figure}
  \includegraphics[width=\columnwidth]{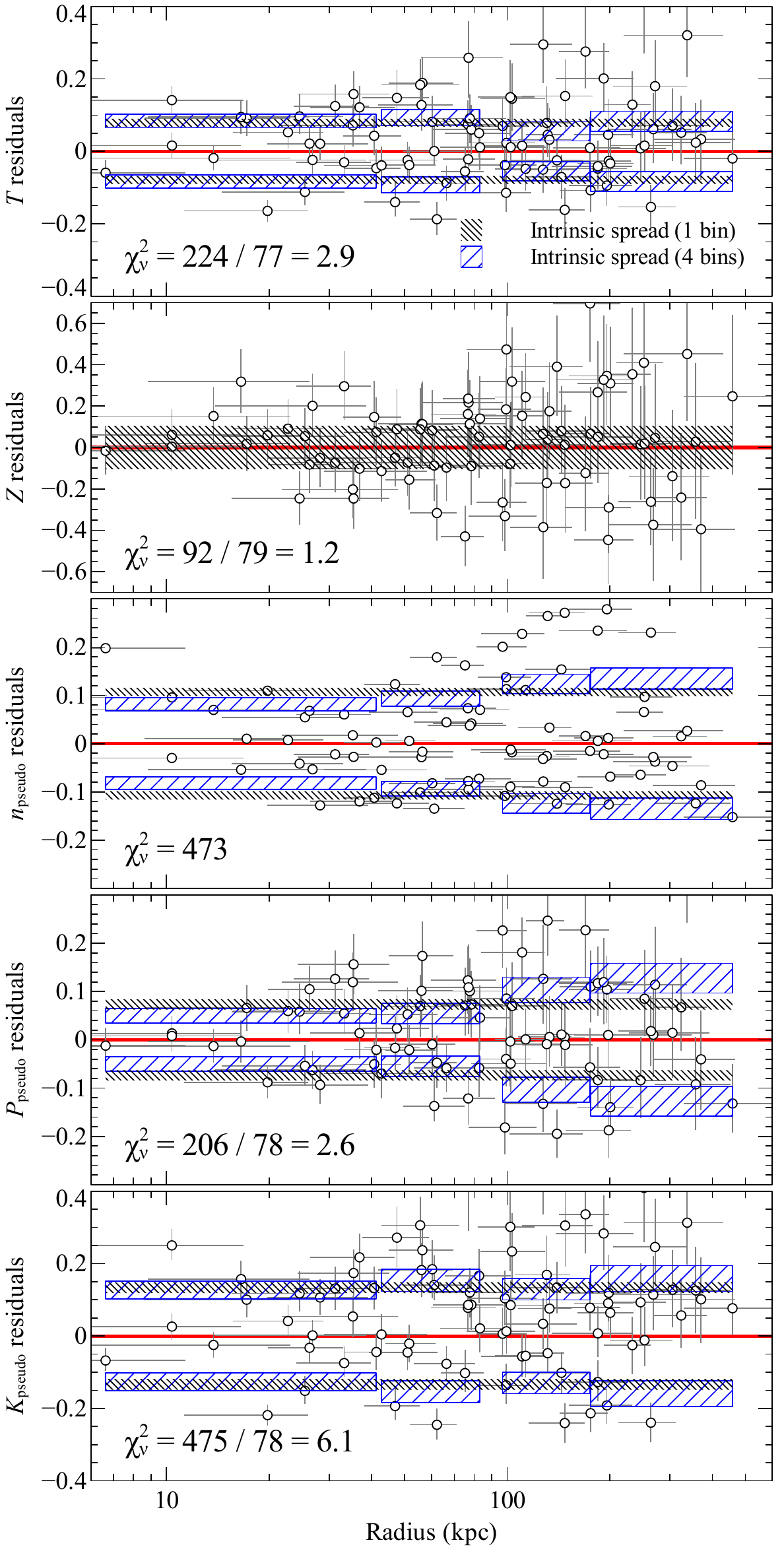}
  \caption{Fractional residuals to the best fitting models in
    Fig.~\ref{fig:map_profiles} with the intrinsic dispersion
    of the points shown. The numeric value show is the
    reduced-$\chi^2$ of the fit. The intrinsic dispersion is
    shown for all points (finely hatched region) and using four radial
    bins containing equal numbers of points (coarse hatched region),
    except for the metallicity plot which shows no significant
    dispersion when split into separate regions. The
    dispersion shows the range of additional
    dispersion which is consistent with Monte Carlo
    realisations of the best fitting models.}
  \label{fig:map_residuals}
\end{figure}

\subsection{Fluctuations in quantities}
The energy in turbulent fluctuations, generated during the growth of a
cluster, is predicted to increase from a few per cent of the thermal
energy density in the centre of a relaxed cluster to tens of per cent
in its outskirts \citep{Vazza09,Vazza11,Lau09}. Unrelaxed objects,
such as those which have previously undergone a merger, have more
turbulent energy at all radii, but particularly in the central region
where it may reach tens of per cent. In addition, the central AGN may
generate motions \citep{Bruggen05,Heinz10}. With current
instrumentation it is possible to directly measure or place limits on
turbulent motions in some cases \citep[see][]{SandersVel13}. However,
the spectra of fluctuations in surface brightness and therefore
density can be measured from images and used as an indirect probe of turbulence
\citep{Churazov12,SandersAWM712,ZhuravlevaFluct14}. By the choice of
energy band for particular temperature ranges, it is possible to
examine pressure fluctuations using surface brightness
\citep{SandersAWM712}. Simulations indicate that density variations
can be used to infer the magnitude of turbulence
\citep{Gaspari14}.

Here, rather than purely relying on surface brightness, we use
spectral information to measure the non-radial dispersion of various
interesting thermodynamic quantities (temperature, density, pressure
and entropy) as a function of radius relative to smooth
models. Measuring these quantities directly provides
useful constraints on turbulence without assuming the results from
simulations.

In Fig.~\ref{fig:map_profiles} are shown the values of the temperature
($T$), metallicity ($Z$), pseudo-density ($n_\mathrm{pseudo}$),
pseudo-pressure ($P_\mathrm{pseudo}$) and pseudo-entropy
($K_\mathrm{pseudo}$) for each bin as a function of the average radius
of the bin.  We exclude two large bins in the outskirts which cover a
very large radial range. Pseudo-density is the square root of the
surface brightness, which is roughly proportional to density times a
line of sight length. We define the pseudo pressure and entropy as
$P_\mathrm{pseudo} = T\,n_\mathrm{pseudo}$ and $K_\mathrm{pseudo} =
T\,n_\mathrm{pseudo}^{-2/3}$, respectively. The solid lines show
radial model fits to the data points, obtained by minimising the
$\chi^2$ statistic. For the temperature profile we fitted the
`universal' temperature profile of \cite{AllenLens01}. A simple linear
function was fitted to the metallicity, a $\beta$ model to the
pseudo-density and pseudo-pressure, and a $\beta$ model with a
negative index to the pseudo-entropy. It can be seen that there are
large amounts of scatter at each radius in some of these quantities,
although the error bars on the metallicities are large. To demonstrate
this, we plot the fractional residuals of each fit to the data in
Fig.~\ref{fig:map_residuals} and quote the $\chi^2$ of the fit.

\begin{table}
  \centering
  \caption{Calculated intrinsic fluctuations in the bin maps over the
    entire radial range, before removing projection effects.
    The correction column shows whether radial symmetry is assumed (No)
    the elliptical \textsc{elfit} surface brightness correction
    has been applied (SB \textsc{elfit}) or the correction from fitting an elliptical
    $\beta$ model to the pressure and density maps (Pressure $\beta$ and Density $\beta$).
    Projection effects are likely to increase these
    values by a factor of $\sim 2$ (Section \ref{sect:projeffects}).}

  \begin{tabular}{lll}
    \hline
    Quantity             & Correction & Dispersion     \\ \hline
    $T$                  & No      & $0.07 \pm 0.01$   \\
    $Z$                  &         & $<0.1$            \\
    $n_\mathrm{pseudo}$  &         & $0.11 \pm 0.01$ \\
    $P_\mathrm{pseudo}$  &         & $0.07 \pm 0.01$   \\ 
    $K_\mathrm{pseudo}$  &         & $0.13 \pm 0.02$   \\ \hline
    $T$                  & SB      & $0.05 \pm 0.01$   \\
    $Z$                  &(\textsc{elfit})  & $<0.1$            \\
    $n_\mathrm{pseudo}$  &         & $0.018 \pm 0.002$ \\
    $P_\mathrm{pseudo}$  &         & $0.05 \pm 0.01$   \\
    $K_\mathrm{pseudo}$  &         & $0.06 \pm 0.01$   \\ \hline
    $T$                  & Pressure& $0.10 \pm 0.01$   \\
    $Z$                  &($\beta$)& $<0.1$            \\
    $n_\mathrm{pseudo}$  &         & $0.069 \pm 0.006$ \\
    $P_\mathrm{pseudo}$  &         & $0.03 \pm 0.01$   \\
    $K_\mathrm{pseudo}$  &         & $0.16 \pm 0.02$   \\ \hline
    $T$                  & Density & $0.10 \pm 0.01$   \\
    $Z$                  &($\beta$)& $<0.1$            \\
    $n_\mathrm{pseudo}$  &         & $0.069 \pm 0.006$ \\
    $P_\mathrm{pseudo}$  &         & $0.03 \pm 0.01$   \\
    $K_\mathrm{pseudo}$  &         & $0.16 \pm 0.02$   \\ \hline
  \end{tabular}
  \label{tab:spreads}
\end{table}

The points will contain some spread due to measurement errors and a
contribution from the intrinsic fluctuations in the cluster at each
radius. We can calculate the intrinsic dispersion, assuming
that the points follow the fitted radial models. We make Monte Carlo
simulations of the quantities assuming that points have the error bars
measured in the data plus an additional fractional
dispersion, added in quadrature. The fraction of simulations
which have a fit statistic better than the real fit statistic is
calculated as a function of the dispersion value. From the
dispersion values where the fraction of better fitting
simulations is 0.50, 0.159 and 0.841 of the total, we calculate the
best fitting dispersion and its uncertainties. These values
are shown for all the points in Fig.~\ref{fig:map_residuals} and in
four radial bins, excluding the metallicity profile.  The average
values for the additional dispersion of the quantities
are given in Table \ref{tab:spreads}. The residual profiles
appear rather flat, although there may be some increase in the scatter
of density and pressure with radius. We do not detect significant
scatter in metallicity.

\subsection{Correcting for surface brightness ellipticity with \textsc{elfit}}
The cluster has an elliptical morphology (Fig.~\ref{fig:resid}). This
will give rise to some of the azimuthal variation of the quantities,
even if the data points were intrinsically smooth. To examine the
effect of the ellipticity, centroid shifts and isophotal twisting, we
again applied the same \textsc{elfit} model with a series of 20
ellipses fitted to the contours in surface brightness at logarithmic
surface brightness contours. We firstly created a corrected radius
map, where for each pixel in the image we calculated the smallest
distances to the nearest two ellipses and calculated the pixel radius
by linearly interpolating between the average radii of those
ellipses. Given this radius image, we gave each bin in the spectral
map a radius from the average radius of the bin pixels in the
corrected radius map. This procedure reduces the effects of isophotal
twisting and edges from the data points.

\begin{figure}
  \includegraphics[width=\columnwidth]{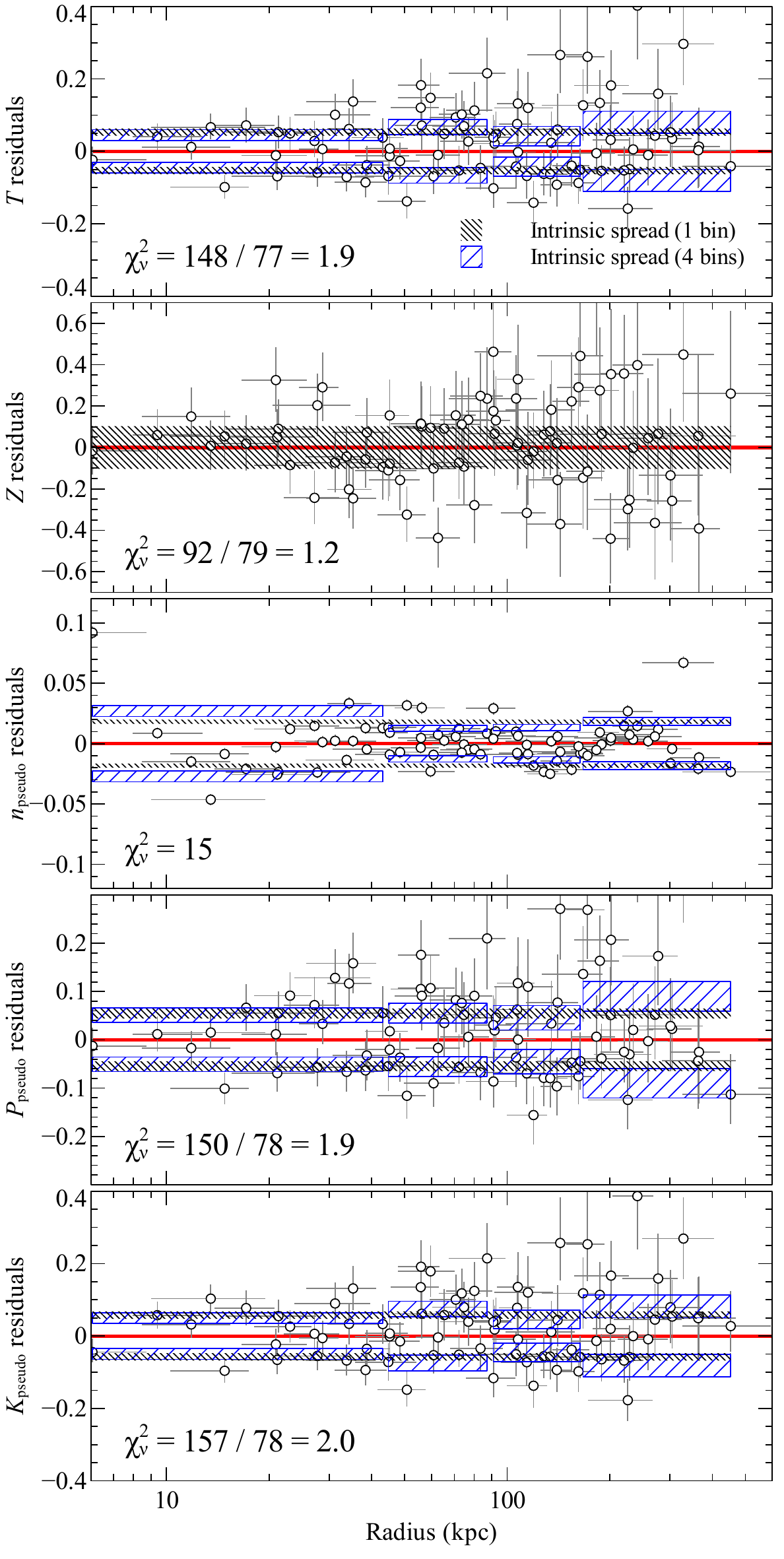}
  \caption{Fractional residuals to the data points after correcting
    the bin radii for the surface brightness ellipticity of the
    cluster on the sky. See Fig.~\ref{fig:map_residuals} for the
    points before correction. Notice the change in axis scale for
    $n_\mathrm{pseudo}$. Large systematic variations were visible in
    the density residuals after correction. By using a double-$\beta$
    model, we removed many of these variations.}
  \label{fig:prof_corrected}
\end{figure}

Fig.~\ref{fig:prof_corrected} shows the residuals from the fits after
correcting the bin radii for elliptical variations. The scatter on all
the quantities except pressure is significantly reduced after this
correction (Table \ref{tab:spreads}). For the density fit, a second
$\beta$ model component was added to the fit to remove significant
radial residuals. The radial profiles of scatter are again rather
flat, but the density plot shows low scatter ($\sim 2$ per cent with
projection) in the middle of the radial range, with larger deviations
in the centre. The central deviations are likely to be due to the AGN
activity.

\subsection{Pressure and density correction with elliptical $\beta$ models}
If the cluster is in hydrostatic equilibrium, the pressure map
  is unaffected by local density or temperature
  perturbations. Therefore the pressure map may better indicate the
  morphology of the cluster potential than the density or surface
  brightness. To investigate whether the scatter in the thermodynamic
  quantities is reduced after correcting for the pressure map, we
fitted the projected pressure map by an elliptical $\beta$ model.  We
minimised the $\chi^2$ between our model and the data assigning an
uncertainty to each pixel of the total for that bin, divided by
$\sqrt{n_\mathrm{pix}}$, where $n_\mathrm{pix}$ is the number of
pixels in the bin. The outermost very large bins were excluded.

We obtain an ellipticity for the pressure map of $0.25 \pm 0.03$
(i.e. a ratio between the minor and major axes is $0.75$) and find the
major axis is inclined $17 \pm 3$ degrees north from the west. The
centre of the model is offset by about 6 arcsec from the nucleus to
the south-east.  The pseudo-density map has a similar ellipticity of
$0.27$, inclined northwards by 13 degrees. By comparing the
ellipticity of the pressure and density, if the cluster is in
hydrostatic equilibrium, it would be possible to estimate the
ellipticity of the dark matter potential.  Fixing the ellipticity and
angle in the pressure fit to the best fitting density values increases
the $\chi^2$ by only 2.4, showing that the difference is not
significant. We do not see any evidence for an elliptical dark matter
potential. However, a gravitational potential is much more symmetric
than the underlying matter density, so measuring this difference is
difficult.

Similarly for the elliptical surface brightness correction, we can
adjust the radii of pixels to correct for the ellipticity of the
pressure distribution. Table \ref{tab:spreads} (under Pressure,
$\beta$) shows that the pressure fluctuations are reduced to around 3
per cent.  The density fluctuations also appear to be reduced compared
to the spherically-symmetric case, but the temperature and entropy
variations are increased.

However, we cannot compare the fluctuations from the pressure $\beta$
modelling to the surface brightness \textsc{elfit} results, because
the $\beta$ correction is relatively crude. It does not, for example,
remove isophotal twisting, centroid shifts or a changing ellipticity
as a function of radius. If we repeat the analysis, fitting the
density map with an elliptical $\beta$ model and doing the radial
correction, we obtain exactly the same results as with the pressure
map (Table \ref{tab:spreads} under Density, $\beta$). This confirms
that the density and pressure ellipticities are very similar.

\subsection{Dependence on bin size}
The magnitude of the fluctuations will have some dependence on the bin
size, as features below the bin size will be smoothed out. We
investigated using bins using a signal to noise ratio of 66 instead of
100 (around 70 per cent of the linear dimension). The computed
additional fluctuations were consistent with each other.

\begin{figure}
  \includegraphics[width=\columnwidth]{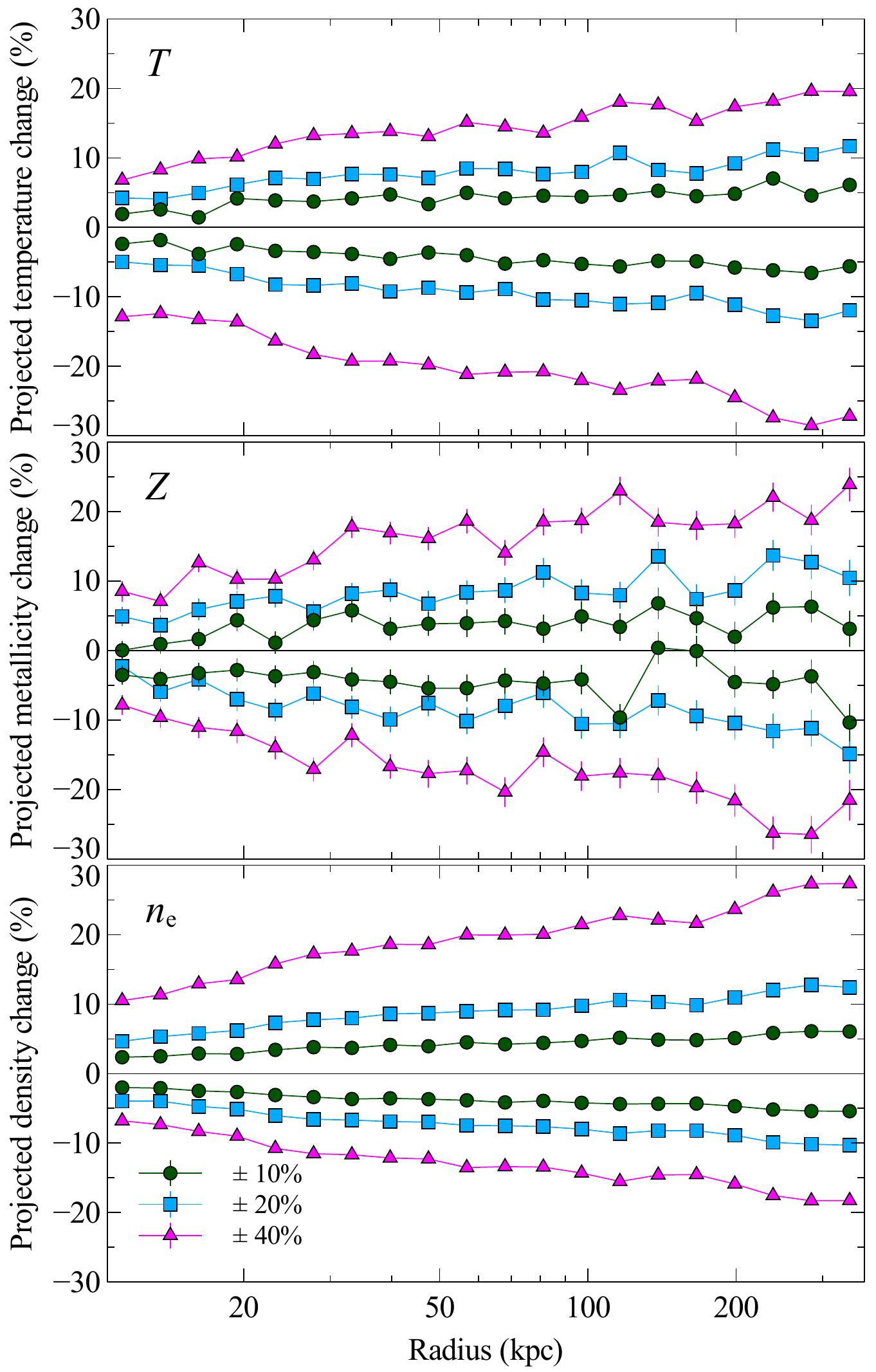}
  \caption{How projection effects affect the measurement of
    fluctuations in quantities as a function of radius. Using
    simulations, fluctuations were inserted in the cluster at a
    particular radius. We assume a bin size equivalent to a signal to
    noise ratio of 100 at a radius and assume the depth of the
    fluctuation along the line of sight is that radius. The values
    show the measured average change in the projected quantity on the
    sky.}
  \label{fig:projfactors}
\end{figure}

\subsection{Projection effects}
\label{sect:projeffects}
The measured dispersions are projected quantities. The
intrinsic dispersion in the cluster itself will be larger, as
fluctuations are smoothed out by the projected emission. We therefore
simulated the effects of projection on fluctuations. From the
deprojected profiles in Section \ref{sect:radial}, we calculated the
projected surface brightness as function of radius on the sky. This
quantity was then converted to the projected bin size giving a signal
to noise ratio of 100, assuming a square bin. For the unperturbed
case, at that radius on the sky, we simulated the spectrum along the
line of sight in that bin, by adding together simulated spectra in
slices out to a cluster radius equivalent to 4.5 arcmin on the sky. We
assumed each slice has constant properties, measured at its midpoint.
From this total spectrum, the best fitting temperature, metallicity
and normalisation was found. The simulation was repeated 160 times to
calculate the mean values and their uncertainties. We then enhanced or
reduced one of the cluster properties (temperature, metallicity or
density) along the line of sight within half the cluster-bin radius
from the cluster midplane. The best fitting quantities were compared
to their baseline values to examine the effect of projection.

Fig.~\ref{fig:projfactors} shows the measured percentage increase or
decrease in temperature, metallicity or density (half the
normalisation change), when the intrinsic cluster emission is
varied. We examine variations in quantities of 10, 20 and 40 per
cent. These profiles show that projection effects are worst in the
centre of the cluster, despite the rising surface brightness
profile. The measured fractional reduction in fluctuation are
reasonably consistent between the different quantities. The strength
of measured fluctuations is reduced to around 25 per cent in the
centre of the cluster, increasing to 50 per cent at a few hundred
kpc. As pressure and entropy are combinations of temperature and
density, they should have similar correction factors.

Note that in reality we do not know the length of fluctuations along
the line of sight, although the radius on the sky is probably a
reasonable approximation. For comparison, if we instead assume the
unlikely case that the fluctuations are concentrated in the midplane
of the cluster and that they are only a bin width thick, we would
obtain fairly uniform projection factors of around 25 per cent in all
quantities.

\section{Discussion}
\subsection{Cooling and heating in the cluster core}
PKS0745 hosts a galaxy cluster where there is evidence for
cooling in many different wavebands. Defining the cooling radius to be
the radius where the mean radiative cooling timescale is less than
$7.7\Gyr$ (the time since $z=1$), it is 115~kpc in PKS0745
(Fig.~\ref{fig:deprojprofs}). Within this radius the cumulative mass
deposition rate calculated from the surface brightness profiles,
taking into account the gravitational work done, is
$730\Msunpyr$. This would be the steady-state cooling rate in the
absence of any form of feedback. Examining the \emph{Chandra} spectrum
inside this radius, we find it consistent with $112 \pm 2 \Msunpyr$
cooling from 6.7 to 0.2 keV temperature (Table
\ref{tab:speccentre}). The RGS instruments on \emph{XMM} are more
suitable for this measurement, although existing observations are
short. However, fixing the model metallicity at the \emph{Chandra}
value, we find consistent results of $270 \pm 90 \Msunpyr$ (Table
\ref{tab:xmmrgs}). The minimum temperature in this case is
$0.5_{-0.2}^{+0.1}$ keV. Splitting the spectrum into temperature bins
(Fig.~\ref{fig:mdots}), the \emph{Chandra} and \emph{XMM} spectra are
largely consistent in showing the spectra are consistent with rates of
$\sim 300 \Msunpyr$ between 4 and 0.5 keV. Below 0.5 keV, the results
are inconsistent, although measuring such cool components with
\emph{Chandra} is very difficult due to effective area, spectral
resolution and calibration uncertainties. In addition, the cluster has
a large Galactic column. Forthcoming \emph{XMM} observations will
enable us to examine the X-ray spectrum in more detail.

Deep observations of several cooling flow clusters have shown that the
picture of a complete cut off in the X-ray temperature distribution is
incorrect. There are many cases now showing clusters with cool X-ray
emitting gas embedded in a hotter medium, associated with emission
line filaments, including Centaurus \citep{SandersRGS08}, Abell 2204,
\citep{SandersA220409}, 2A\,0335+096 \citep{Sanders2A033509}, Abell
2052 \citep{dePlaa10}, M87 \citep{Werner10}, S\'ersic 159
\citep{Werner11}, Abell 262 and Abell 3581 \citep{SandersRGS10} and in
several elliptical galaxies \citep{Werner14}.

With \emph{Chandra} we are able to resolve the coolest X-ray emitting
material in the core of the cluster (Fig.~\ref{fig:nuclimg}), finding
it to consist of a 10-kpc-wide 2.5-keV-average-temperature blob of
material to the west of the cluster core. Radial profiles
(Fig.~\ref{fig:deprojprofs}) confirm the central temperature. These give
a central radiative cooling time and entropy of around $500 \Myr$ and
$10 \keVcmsq$, respectively.

The central galaxy of the cluster is observed to be blue and to
contain ionised line-emitting gas extended over around 7~kpc from the
core of the central galaxy \citep{Fabian85,McNamara92}.  This coolest
material is coincident with the some of the brightest parts of the
ionised line-emitting nebula (Fig.~\ref{fig:nuclimg} bottom panel).
Vibrationally-excited molecular Hydrogen was observed in the
PKS0745 using \emph{HST} \citep{Donahue00}, with a morphology
similar to the ionised line-emitting material. Cold molecular gas, as
seen using CO emission, has been observed from the cluster
\citep{SalomeCombes03}. \cite{ODea08} and \cite{Hoffer12} measured a
star formation rate of $17-18\Msunpyr$ using \emph{Spitzer} infrared
photometry. With \emph{Spitzer} spectroscopy \cite{Donahue11} obtained
a value of $11 \Msunpyr$. Much higher rates of $130-240 \Msunpyr$ were
obtained using ultraviolet \emph{XMM} optical monitor observations
\citep{Hicks05}. However, this UV emission may have a different origin
from star formation.

In PKS0745 we are clearly observing a region around the core
where there are a wide range of different temperature phases. The
X-ray spectra are consistent with several hundred $\Msunpyr$ cooling
out of the X-ray band. Close feedback could be operating, reducing the
star formation rate to a few percent of the X-ray cooling
rate. However, \cite{Fabian11} propose that the emission line
filaments in the centres of clusters are powered by secondary
electrons generated by the surrounding hot gas, explaining their
peculiar low excitation spectrum. Here, the gas cooling accretion
rate, $\dot{M}$, is related to the luminosity of the cool or cold gas,
in units of $10^{43} \ergps$, $L_{43}$ and temperature in $10^7$~K,
$T_{7}$, by $\dot{M} \approx 70 \, L_{43} T_7^{-1} \Msunpyr$. The
H$\alpha$ emission is $3 \times 10^{42} \ergps$ \citep{Heckman89}, but
the total luminosity should be 10--20 times larger. The luminosity is
therefore consistent with rates of several hundred $\Msunpyr$
cooling. The X-ray gas could cool to 0.5 to 1 keV and then merge with
the cool gas, producing the bright H$\alpha$ emission in this object.

One of the notable aspects of the cooling in this object is that the
coolest X-ray emitting gas and emission line nebula is offset from the
nucleus. Such an offset has been seen before in several other
well-known clusters, including Abell 1991, Abell 3444 and Ophiuchus
\citep{Hamer12} and Abell 1795 \citep{Crawford179505}. \cite{Hamer12}
suggest that offsets occur in 2 to 3 per cent of systems and present
some possible mechanisms for the gas-galaxy displacement, including
sloshing of the gas in the potential well, which seems a likely
candidate here given the strong cold front in this cluster.

Although there is apparently cooling taking place, there are at least
two central X-ray surface brightness depressions indicating AGN
feedback. These are likely to be cavities filled with bubbles of radio
emitting plasma which are displacing the intracluster medium, as seen
in many other clusters \citep{McNamaraNulsen12}. They are one of the
means by which AGN appear to be able to inject energy which is lost by
cooling \citep{Fabian12}.  We do not observe the corresponding radio
emission here, but the central radio source extends in the direction
of the cavities (Fig.~\ref{fig:nuclimg}). In PKS0745, these cavities
are around 5 and 3 arcsec in radius (or 17 and 9.8 kpc). The central
total thermal pressure is around $0.6 \keV \pcmcu$. Therefore, the
total bubble enthalpy, assuming $4PV$, is $3 \times 10^{60} \erg$
(Fig.~\ref{fig:deprojprofs}). If the bubble rises at the sound speed
(assuming 4 keV material), this would give a timescale of around
20~Myr using a radius of 17 kpc. The heating power of these bubbles is
therefore around $5 \times 10^{45} \ergps$. Therefore, the energetics
of these bubbles would be sufficient to offset cooling in the core of
this cluster, although this is an order of magnitude estimate given
the messy central morphology. As the star formation is a few
  per cent of the mass deposition rate, the feedback is almost
  complete, or the cooling energy goes to power the nebula as
  suggested above.

It is interesting to compare PKS0745 to the Phoenix cluster
\citep{McDonald12}, the most extreme star forming cluster currently
known, which lies at a redshift of 0.596 and has the highest-known
cluster X-ray luminosity ($8.2 \times 10^{45} \ergps$ between 2 and 10
keV). Phoenix has a classical mass deposition rate of $\sim 1900
\Msunpyr$ \citep{McDonald13SPT} and the central galaxy has a current
star formation rate of $\sim 800 \Msunpyr$
\citep{McDonaldStarPheonix13}. The total H$\alpha$ luminosity from the
Phoenix cluster is $8 \times 10^{43} \ergps$ \citep{McDonald14},
compared to $3 \times 10^{42} \ergps$ for PKS0745. The H$_2$
molecular gas mass in Phoenix is $2.2 \times 10^{10} \Msun$, compared
to $4 \times 10^{9} \Msun$ yr. Examining the ratios of quantities
between PKS0745 and Phoenix, the classical mass deposition rate
ratio is $\sim 0.4$, the molecular gas mass ratio is $\sim 0.2$, the
H$\alpha$ luminosity ratio is $\sim 0.04$ and the star formation rate
ratio is $\sim 0.015$. PKS0745 has a comparable mass deposition rate
to Phoenix and its molecular mass ratio is consistent, but the
resulting star formation rate and H$\alpha$ luminosity is much
lower. Phoenix is clearly converting molecular material to stars much
more rapidly than PKS0745. A major difference between the two
objects is the highly luminous obscured central AGN in Phoenix.
In addition, molecular material can survive for Gyr timescales
in clusters before it
forms stars \citep[e.g.][]{Canning14}. It is therefore surprising that the
Phoenix cluster is forming stars so rapidly that it will exhaust its molecular
material in 30 Myr, unless replenished \citep{McDonald14}.

\subsection{Cluster profiles}
In Section \ref{sect:radial} we use a new code to extract cluster
thermodynamic properties from surface brightness profiles. The results
from this code match those from conventional spectral fitting. The
advantages of this technique over spectral methods is that it does not
require spectral extraction and fitting, it can examine smaller
spatial bins, it gives confidence regions on the mass profile of the
cluster and it can operate on datasets containing only a few hundred
counts.

We examined the gas properties across the four edges in surface
brightness to the east and west of the cluster core (Section
\ref{sect:edges}; Fig.~\ref{fig:jumpprofile}). The profiles are
consistent with continuous pressure across the edges. This indicates
that they are cold fronts \citep{MarkevitchCFShock07}, i.e. contact
discontinuities.  There is some indication of a jump in pressure
across the innermost western edge, which could be a weak shock
associated with AGN feedback, as seen for example in Perseus
\citep{FabianPer03}, Virgo \citep{FormanM8707} and A2052
\citep{Blanton11}. The swirling morphology in surface brightness and
temperature (Figs. \ref{fig:resid} and \ref{fig:centremaps}) is
similar to that seen in simulations where the cluster gas is sloshing
in the potential well.  However, the 2D map shows that a simple cold
front is probably too simple a description in this case, with high and
low pressures in the northern and southern halves of the edge,
respectively. This may contribute to the finite width of the western
edges, although this could be because they are not perfectly in the
plane of the sky or it could be due to a mismatch between the shape of
our extraction regions and the edge. The kinked shapes of the cold
fronts to the east of the core suggest that Kelvin-Helmholtz
instabilities are operating there \citep{Roediger12}.

When examining the radial profiles for the whole cluster, the King
mass model gives a better fit to the data than the NFW model. The NFW
model appears to incorrectly model the gravitational acceleration
($g$) in the centre of the cluster. The King and Arb4 models suggest
that $g$ declines to small central values. If the NFW model is fitted
to the outskirts, its $g$ values agree with the King and Arb4 models
there. The NFW model parameters for the full data range gives a small
concentration of 3.9 and a mass of $M_{200} = 1.3 \times 10^{15}
\Msun$. Fitting the outer profile gives a large concentration of 9.2
and $M_{200} = 3.6 \times 10^{14} \Msun$. These compare to values of
$5.3^{+0.5}_{-0.9}$ and $9.8^{+1.9}_{-1.0} \times 10^{14} \Msun$
calculated from a \emph{Suzaku} analysis to the virial radius and
beyond \citep{Walker12}. Our analysis is only able to probe to radii
of around 500~kpc and has difficulties fitting the data inside 40~kpc,
which is likely biasing the obtained parameters. Our agreement of the
mass profile over our outer range with \cite{Walker12} is very good
(lower left panel of Fig.~\ref{fig:deprojprofs}).

With the King and Arb4 models our method finds low values of $g$
within 40 kpc. The projection method self consistently includes the
dark matter and gas mass, but does not include the central galaxy or
its supermassive black hole. However, including the galaxy or black
hole could only increase the value in $g$. Our modelling prefers low
values of $g$, where there would be no significant dark matter in the
central regions. $g$, however, is essentially driven by the pressure
gradient. The very flat pressure profile seen both spectrally and
using \textsc{mbproj} would imply very low central mass densities in
the cluster given hydrostatic equilibrium. We have confirmed that the
spectral pressure profiles agree with the values of $g$ we obtain with
\textsc{mbproj}.  However, the assumptions of the model may be
broken. There could be a strong lack of spherical symmetry, the ICM
may not be in hydrostatic equilibrium or there could be strong
non-thermal contributions to the central pressure.

There is structure within the inner part of the cluster, indicating
significant non-spherical variation. Within 40~kpc are the strongest
signs of cooling and the cavities within the X-ray emission
(Fig. \ref{fig:resid}).  The two cold fronts seen to the west and two
to the east of the cluster core could indicate that the gas is
sloshing within the potential well. This may disturb hydrostatic
equilibrium giving velocities of a few hundred $\kmps$
\citep{Ascasibar06}. The inner cold fronts are on the same scales as
where the mass model becomes unrealistic. In addition, the feedback
from the central AGN could induce similar velocities in the ICM
\citep{Bruggen05,Heinz10}. Other non-thermal sources of pressure may
include cosmic rays and magnetic fields and may be associated with the
central radio source. We note that there are indications of a pressure
jump at the location of the inner western cold front, suggesting the
cluster is not in hydrostatic equilibrium there. It is possible that
there is a weak shock surrounding the central cavities. The
  central thermal pressure obtained without assuming hydrostatic
  equilibrium is $8 \times 10^{-10} \erg\pcmcu$. If we calculate the
  central pressure obtained assuming the NFW mass model fitted beyond
  0.5 arcmin radius (Fig.~\ref{fig:deprojprofs}), this yields a
  central pressure of $1.6 \times 10^{-9} \erg\pcmcu$. Therefore, if
  the flat pressure profile is due to a non-thermal source of
  pressure, it has the same magnitude as the thermal pressure.  The
  implied magnetic field strength is $140 \mu$G if the non-thermal
  pressure is magnetic. Taking a central density of $0.1\pcmcu$ and a
  path length-length of 10 kpc, gives a rotation measure of $10^5$,
  implying that the radio source should be completely depolarized on kpc
  scales. Indeed, \cite{BaumODea91} found that the source is strongly
  depolarized. 

\cite{Allen96} used strong gravitational lensing to measure the
projected mass in the centre of the cluster. They obtained a projected
mass of $3 \times 10^{13} \Msun$ within the critical radius of 45.9
kpc, or $2.5 \times 10^{13} \Msun$ with a more sophisticated
analysis. These values are for a cosmology using $H_0 = 50 \kmpspMpc$,
$\Omega=1$ and $\Lambda=0$. In our cosmology the projected mass within
34.4 kpc radius is $1.8 \times 10^{13} \Msun$, if we scale the
  value from the more sophisticated analysis using their equation
  3. Taking our NFW and King best fitting models and projecting the
mass inside this radius on the sky, we obtain $1.2 \times 10^{13}$ and
$8.1 \times 10^{12} \Msun$, respectively. These values confirm that
our models are missing a substantial mass in the central region. This
is likely because of the low $g$ values we obtain and the
lack of a central galaxy in our mass modelling.

\subsection{Sloshing energetics}
If the cold fronts are caused by the sloshing of the gas in the
potential well, a large amount of energy may be stored in this
motion. To estimate the potential energy, we take the radial electron density
profile in the western sector and compare it to the profile for the
complete sector (both profiles are shown in
Fig. \ref{fig:jumpprofile}). The mass of extra material in the west,
for a shell at radius $r$ with width $\delta r$ is
\begin{equation}
  \delta M(r) =  (n_\mathrm{e,west} - n_\mathrm{e,complete}) \,
  \mu m_\mathrm{H} Y \, \Omega r^2 \delta r ,
\end{equation}
where the electron densities in the complete and western sector are
$n_\mathrm{e,complete}$ and $n_\mathrm{e,west}$, respectively, $\mu$
is the mean molecular weight, $m_\mathrm{H}$ is the mass of a Hydrogen
atom and $Y$ is a factor to convert from electron to total number
density. We assume that the sloshed region occupies a solid angle
$\Omega \sim 1.8$ (a cone with an opening angle of 90 degrees). The
derived total excess gas mass in the western direction is $1 \times
10^{11} \Msun$ between 14 and 200 kpc radius (we note that the excess
mass outside this radius is larger, however).

To estimate the potential energy for a particular shell and radius, we
take its density and find the radius in the complete cluster profile
which has the same density. This radial shift is then converted to a
difference in cluster potential using the best fitting King model
(NFW results are consistent),
assuming it has simply shifted between these radii without a change in
density. The shell excess mass ($\delta M$) and the potential
difference are multiplied to calculate an estimate of the potential
energy of the sloshing for that shell.  By adding the energy of the
shells between radii of 14 and 200 kpc, we estimate that the total
potential energy of the sloshed gas in the western sector is $3 \times
10^{59} \erg$. In the eastern sector there are also cold fronts, so a
similar amount of energy may be stored there. Unfortunately a similar
analysis is difficult there because  the
eastern sector is not centred on the cluster centre due to the cold
front morphology.

The lifetimes of cold fronts in simulations are of the order of Gyr
\citep{Ascasibar06,ZuHone10}. If the gas sloshing potential energy
could be converted to heat over this period, it only represents a
heating source of $\sim 10^{43} \ergps$, much weaker than the central
AGN. However, the dark matter potential energy contribution could be
larger.

\cite{ZuHone10} suggest that sloshing can contribute to the heating in
clusters, although the mechanism in their simulations is that hot gas
and cooler material are brought together and mixed. In addition the
cluster core can be expanded, reducing the radiative cooling. The
effectiveness of the heat flux is reduced if the ICM is
viscous. However, recent simulations by \cite{ZuHone14} including
Braginskii viscosity or Spitzer viscosity and magnetic fields, find
that these effects can significantly change the development of
instabilities and turbulence, which is likely to affect the amount of
hot and cold gas mixing.

\subsection{Thermodynamic fluctuations}
In the central region we see significant deviations from spherical
symmetry in projected surface brightness ($\sim 20$ per cent, or 10
per cent in density), 20 per cent in temperature and 15 per cent in
pressure (Section \ref{sect:centmaps}).

In Section \ref{sect:largemaps} we measured the additional fractional
variations (assuming Gaussian deviations) required to the data points
consistent with the smooth fitted profiles. Projected fluctuations of
7 per cent in temperature, less than 10 in metallicity, 11 per cent
in density, 7 per cent in pressure and 13 per cent in entropy
were measured (Fig.~\ref{fig:prof_corrected} and Table
\ref{tab:spreads}).  As in the Perseus cluster \citep{SandersPer04},
we find that the scatter in temperature and density conspires to give
a smaller fractional scatter in pressure than would be expected if
they were uncorrelated.  Projection effects mean that the intrinsic
variations are larger. The intrinsic variations of temperature,
metallicity and density are around 4 times larger in the centre
(Fig.~\ref{fig:projfactors}) to twice as large at 300~kpc.

However, Fig.~\ref{fig:resid} shows that much of the surface
brightness (or pseudo-density) variation comes from the non-spherical
nature of the cluster and the cold fronts. We corrected for this by
moving the points radially, by comparison to the positions of ellipses
fitted to surface brightness contours using \textsc{elfit}. This
reduces the dispersions in temperature and density to 5 and 2
per cent, respectively. The variations in pressure are reduced to 5
per cent. Some of the remaining density variation appears to be
systematic (likely the spiral morphology remaining in
Fig.~\ref{fig:resid}). After this correction we observe similar
variations in temperature and pressure, but smaller density
variations. Projection effects likely increase the 3D density
variation to around 4 per cent.

A 4 per cent density variation was also inferred in AWM\,7
\citep{SandersAWM712}, where we compared observed fluctuations to
models including a 3D turbulent spectrum. In addition, there were
regions in that cluster with only 2 per cent inferred fluctuations. In
the more disturbed Coma cluster, \cite{Churazov12} found 7 to 10 per
cent density variations, reducing to 5 per cent on 30 kpc scales.

The pressure variation does not reduce as strongly as the density
variation after correcting for the surface brightness with
\textsc{elfit}. This suggests that the pressure does not closely
follow the surface brightness variations. It is likely closer to a
hydrostatic atmosphere in nature with additional pressure
fluctuations. Correcting for the overall ellipticity of the pressure
or density maps with an elliptical $\beta$ model reduces the pressure
fluctuations from 3 per cent (where it is 7 per cent in the spherical
case or 5 per cent in the \textsc{elfit}-corrected case), supporting
the idea of a hydrostatic atmosphere. Therefore the temperature and
density show structure beyond simple ellipticity (such as
centroid-shifts, radial ellipticity variation and twisting isophotes)
which do not correspond with those seen in the pressure distribution.
The density, temperature and entropy are strongly correlated with the
more structured surface brightness structure. It should be noted that
the size of the bins examined increases as a function of radius. If
the size of the structures being examined does not increase in the
same way, we will differentially smooth features as a function of
radius.

In this cluster we do not observe significant non-statistical scatter
in projected metallicity, after correcting the bin radii for the
elliptical morphology, although 20 per cent deprojected scatter is
allowed by the data. In several clusters, high metallicity blobs of
material have been found embedded in lower metallicity material
(e.g. Perseus, \citealt{SandersPer07}; Abell 85,
\citealt{Durret05}; NGC 4636, \citealt{OSullivan05}; Abell 2204,
\citealt{SandersA220409} and S\'ersic 159, \citealt{Werner11}).

Simulations by \cite{Gaspari14} suggest that the fractional density
variation should be close to the 1D Mach number and so it is
interesting to compare our results to theirs. However, we are limited
in our comparison because we measure projected fluctuations and
currently do not measure the size of projections as a function of
length scale.

The results we obtain depend strongly on whether we correct for
asymmetries in the cluster. If we use the assumption of radial
symmetry, we obtain density variations of around 20 per cent (after
correcting by a factor of 2 for projection). Temperature variations
are around 20 per cent, pressure fluctuations 12 per cent and entropy
variations 30 per cent (Table \ref{tab:spreads}). The density
variations imply 1D Mach numbers of around 20 per cent ($\sim 300
\kmps$ at 8 keV). In this lower Mach number regime, \cite{Gaspari14}
suggest that density and pressure variations should be
similar. Entropy variations should be larger than these and the
pressure smaller. This is what we find in our data assuming spherical
symmetry.

If we instead correct our data for ellipticity in the surface
brightness, removing much of the sloshing structure using the
\textsc{elfit} model, the density variations become much smaller (4
per cent). This implies velocities of the order of only $70 \kmps$. In
this analysis the entropy fluctuations are reduced to be the same as
the pressure fluctuations, which differs from the \cite{Gaspari14}
results.  Correcting for pure overall pressure or density ellipticity
reduces the density variations from the spherical case to around 14
per cent. The entropy fluctuations, however, are not reduced.

Therefore comparison with theory depends on what is considered to be
the underlying model on which the fluctuations are measured. If the
sloshing morphology is seen as part of the turbulent cascade, then it
should remain in the analysis. It, however, dominates the
calculation of the density, temperature and entropy
fluctuations.

\subsection{Application of \textsc{mbproj} to the low count regime}
\begin{figure}
  \includegraphics[width=\columnwidth]{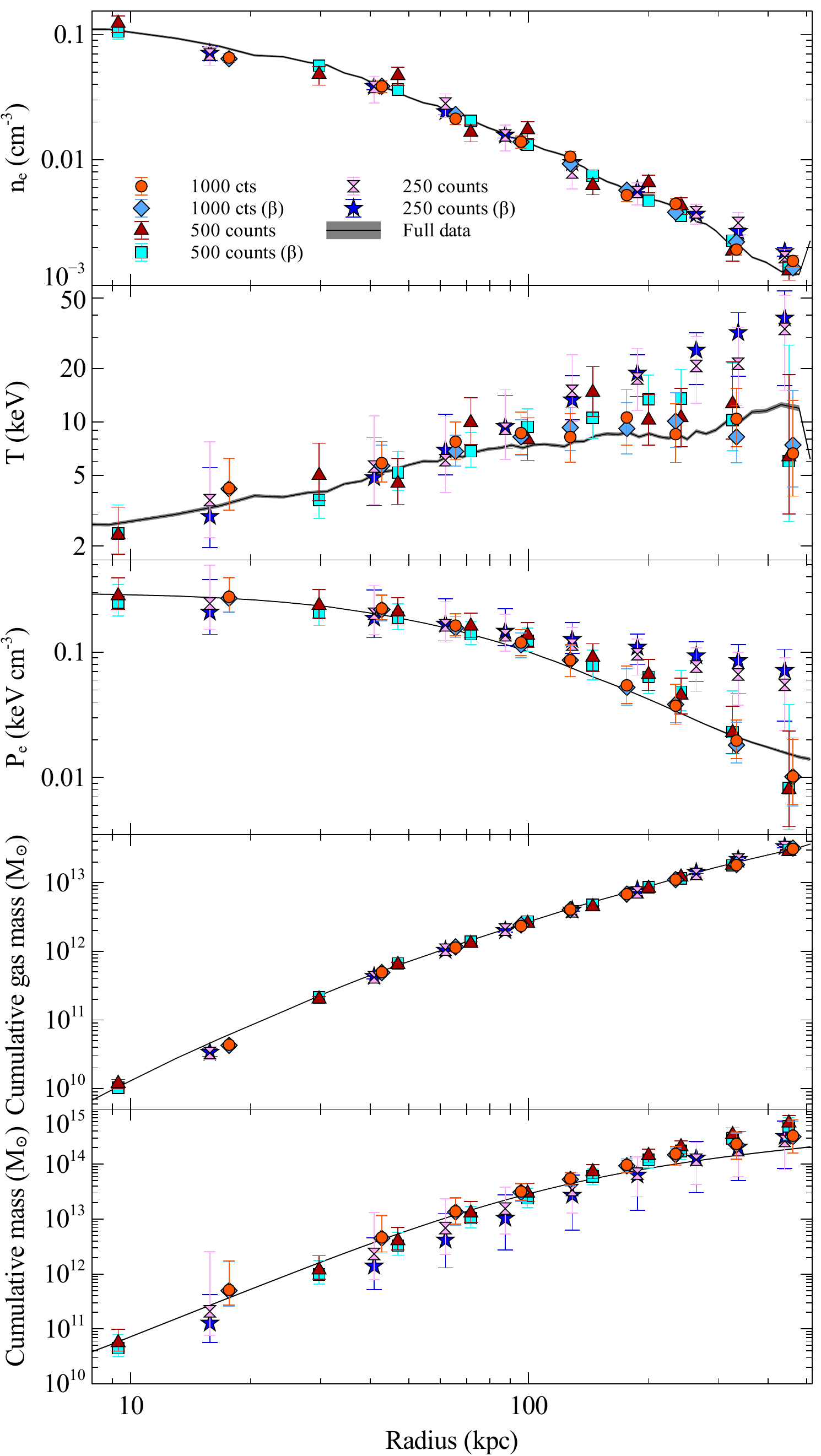}
  \caption{Results from \textsc{mbproj} King model analyses of
    realisations of PKS0745 surface brightness profiles with
    short exposure times. Shown are results for clusters with 1000,
    500 and 250 counts. For each realisation, we show the results
    assuming independent densities in each bin and assuming a $\beta$
    model.}
  \label{fig:reduce}
\end{figure}

In this paper we applied the \textsc{mbproj} method to very good
quality data. However, the technique can operate in the low count
regime, as demonstrated in Fig.~\ref{fig:reduce}, which shows results
from analysing three realisations of the PKS0745 surface brightness
profiles, with 1000, 500 and 250 cluster counts in total. We assume zero
background in these simulations. The method is able to reproduce the
full density profile and gas mass profile in each case. Reasonable
temperature and pressure profiles are obtained with 1000 and 500 count
profiles. The total mass profiles are also consistent with the full
case, although the uncertainties are large in fractional terms ($\sim
50$ per cent for the 1000 count dataset). In this analysis we assume
flat priors on the core radius (between $1 \kpc$ and $2.5 \Mpc$) and
on the velocity dispersion (between 10 and $2000 \kmps$). It is likely
that improved priors on the mass model will reduce these
uncertainties, although the outer pressure is one of the main
uncertainties.

This simple simulation does not include the effects of background or
the telescope point spread function (PSF) and only examines a single
realisation of a single object, but it shows that the modelling method
is likely to be useful in examining data from current and future X-ray
cluster surveys (e.g. using \emph{eROSITA};
\citealt{Predehl10,Merloni12}). Such a technique would extract the
maximum available information from each object and could be a useful
improvement over scaling relations.  The analysis of better
simulations and existing cluster surveys will test the usefulness of
the technique for surveys. Indeed, the technique may work better in
lower mass objects as it is difficult to measure temperatures using
three bands at temperatures of $\sim 10$~keV.

\section{Conclusions}
We present a new technique, \textsc{mbproj}, for the analysis of
thermodynamic cluster profiles without the use of spectral fitting. It
assumes hydrostatic equilibrium, spherical symmetry and a dark matter
mass model and uses MCMC to deduce the uncertainties on the various
thermodynamic quantities. In this cluster, the low obtained
gravitational acceleration suggests that the assumptions of
hydrostatic equilibrium or spherical symmetry are invalid or that
there are additional non-thermal sources of pressure. The code
works in the low count regime, suggesting it will be very useful for
the analysis of cluster survey data.

The analysis of new \emph{Chandra} and \emph{XMM} observations of the
PKS0745 galaxy cluster suggest that there is at least a factor of 20
in X-ray gas temperature in this object (from 10 keV to at least 0.5
keV). There is no sharp cut-off in X-ray temperature. It appears,
despite the evidence for feedback in the form of central cavities,
that cooling of the intracluster medium is occurring at the rate of a
few hundred solar masses per year. As found in several other clusters,
the coolest material X-ray emitting material and line emitting nebula
is offset from the central AGN.

There are two sets of cold front to either side of the nucleus. This
could be suggesting sloshing of the gas in the potential well, which
would also explain the offset of the coolest material. We investigate
the azimuthal variation of projected thermodynamic quantities in the
cluster. If the cluster were spherical, we find projected variations
of 7 per cent in temperature and pressure and 11 per cent in
density. The entropy variation is 13 per cent. If we correct for the
shifting isophotes in the cluster, the density variation reduces to 2
per cent and the temperature and entropy variation to 5 and 6 per
cent, respectively. Projection effects are likely to increase these by
around a factor of 2. The magnitude of the fluctuations depends
strongly on how the signal from the underlying cluster is subtracted
and so should be considered when comparing to theoretically predicted
values.

\section*{Acknowledgements}
ACF, HRR and SAW acknowledge support from ERC Advanced Grant FEEDBACK.
JHL is supported by NASA through the Einstein Fellowship Program,
grant number PF2-130094.  GBT acknowledges the support provided by the
National Aeronautics and Space Administration through Chandra Award
Numbers G01-12156X and GO4-15121X issued by the Chandra X-ray
Observatory Center, which is operated by the Smithsonian Astrophysical
Observatory for and on behalf of the National Aeronautics Space
Administration under contract NAS8-03060.  The scientific results
reported in this article are based on observations made by the
\emph{Chandra X-ray Observatory}. This article contains results based
on observations obtained with \emph{XMM-Newton}, an ESA science
mission with instruments and contributions directly funded by ESA
Member States and NASA. We thank an anonymous referee for providing a
helpful report.

\bibliographystyle{mn2e}
\small
\bibliography{refs}

\end{document}